# An Intelligent System for Multi-topic Social Spam Detection in Microblogging

Bilal Abu-Salih[1] {b.busalih@ju.edu.jo }, Dana Al Qudah[1], Malak Al-Hassan[1], Seyed Mohssen Ghafari[2],

Tomayess Issa[3], Ibrahim Aljarah[1], Amin Beheshti[2], Sulaiman Alqahtani[2]

[1] The University of Jordan

[2] Macquarie University

[3] Curtin University

**Abstract:** The communication revolution has perpetually reshaped the means through which people send and receive information. Social media is an important pillar of this revolution and has brought profound changes to various aspects of our lives. However, the open environment and popularity of these platforms inaugurate windows of opportunities for various cyber threats, thus social networks have become a fertile venue for spammers and other illegitimate users to execute their malicious activities. These activities include phishing hot and trendy topics and posting a wide range of contents in many topics. Hence, it is crucial to continuously introduce new techniques and approaches to detect and stop this category of users. This paper proposes a novel and effective approach to detect social spammers. An investigation into several attributes to measure topic-dependent and topic-independent users' behaviours on Twitter is carried out. The experiments of this study are undertaken on various machine learning classifiers. The performance of these classifiers are compared and their effectiveness is measured via a number of robust evaluation measures. Further, the proposed approach is benchmarked against state-of-the-art social spam and anomalous detection techniques. These experiments report the effectiveness and utility of the proposed approach and embedded modules.

## 1. Introduction

The rise of Online Social Networks (OSNs) has led to reinforce values of freedom of expression, communication, and breaking the monopoly of information. Therefore, these virtual platforms have been used to circulate information, support movements of rejections and protests, expose corrupt practices and other various activities. Despite the plethora of benefits brought by the continuous use of these platforms, the lack of 'gatekeepers' has opened wide the door to carrying out fraud, defamation and speeding of rumours, tarnishing the reputation of organisations and individuals, to all other types of false information that have permeated these platforms. This poses significant challenges as various malicious activities have degraded the quality of experience obtained by the members of these virtual communities [1]. Social spam is commonly referred to as nasty activities or unsolicited and low-quality content that spread over the OSNs. Examples of such activities include profile cloning, social phishing, fake reviews, bulk submissions, hashtags hijacking, clickjacking, sending malicious links, etc. Social spammers are those who inject spam and practice such activities over the OSNs [2]. Despite the ongoing efforts to cease such activities and keep the platforms clean, the social spam phenomenon continues to rise; almost 47 percent of respondents to a recent survey indicate perceiving more spam into their social media feeds [3].

To gain credibility, this category of users attempts to establish dialogues with people of dissimilar interests by attempting to allure non-spammers into befriending them or by carrying out hashtag hijacking or bulk messaging of a specific topic or a wide range of topics [1, 4]. Topics of interest are particular areas of an individual's work, expertise, or specialisation within the scope of subject-matter knowledge such as science, politics, sports, education etc. [5-7]. This implies that a user's credibility in OSNs is topic-driven; users can be trustworthy and reliable in one or few topics, yet this does not apply to other topics of interest. In OSNs, we argue that there is an inverse relationship between the number of topics the user is interested in and the user's topic-based (domain-based) credibility. This argument is justified based on the following facts: (i) there is no well-informed legitimate user who has the intellectual capacity to publish contents in all topics [8]; (ii) users who publish contents on various topics of interest do not convey to other users the particular topic that they are interested in. This is evident as topics of interest can be intuitively deduced from content of users who commonly post wide-ranging content in one or few topics; (iii) it is likely that this user is a spammer or anomalous; this illegitimate category of users tends to post tweets about numerous topics [9, 10]. This could end up with tweets being posted on all topics which do not convey a legitimate user's behaviour. Therefore, distinguishing users in a set of topics is a significant aspect.

The existing social spam detection approaches are generic in terms of topic extraction, and they passively extract features based merely on users' data and metadata. In particular, the current approaches incorporate a probabilistic generative model, namely the Latent Dirichlet Allocation (LDA) [11] to distil users topics of interest from their textual data. LDA and similar other statistical models fail to capture high-level topics and to consider and integrate the semantic relationships between terms in textual data. Also, they are inadequate to extract correct topics from short textual messages such as tweets [12, 13]. Another drawback of the existing works is that they neglect to extract sentiments of the tweets' conversations (tweets replies), thereby unable to listen to the subjective impressions of users' followers towards tweets' contents. On the other hand, our study proposes a novel and effective multi-topic social spammers detection model for microblogging. To distinguish users' topics of interest, we incorporate the idea of distinguishing/discriminating which was developed in the Information Retrieval (IR) domain through applying $tf.idf$ formula [14]. "The intuition was that a query term which occurs in many documents is not a good discriminator" [15]. This implies that a term that occurs in many documents decreases its weight in general as this term does not show which document the user is interested in [16]. This heuristic aspect is incorporated into our model as an important driver to extract features from the users' contents based on their topics of interest. The users' topics of interest have been investigated and analysed using two approaches, namely: IBM Watson – Natural Language Understanding (NLU) API [17]) and a developed topic discovery model (based on 20 Newsgroups dataset [18]). IBM Watson NLU, formally known as Alchemy API, involves analysing textual content and extract semantic features incorporating machine learning, semantic web technologies, and linked open data. These semantic features include entities, emotion, keywords, relations, categories/taxonomies (currently 23 topics), sentiment analysis, etc. IBM Watson NLU provides both topic-independent and customized topic analysis that can be implemented using their Knowledge Studio. This designated API is used both to extract users' domains of knowledge as well as to infer sentiments of followers' replies as it will be discussed later.

Further, we developed a topic classification module based on the 20 Newsgroups dataset and by implementing a number of machine learning classification models. The set of topics extracted by both models are mapped and aggregated, thereby inferring a unified model for topic discovery that will be used further in the proposed spammer detection model. In particular, we investigate several attributes to measure topic-dependent and topic-independent users' behaviours in Twitter. A total of 18 key features is attained and extracted from contents and user analysis. The experiment of this study was carried out on manually labelled Twitter datasets. The list of user_ids used for data collection is stemmed from two popular corpora: (i) topically anomalous dataset [19]. This graph is chosen since it includes the list of Twitter users who had less than 5,000 friends. This threshold was initially established to discover anomalous and other illegitimate users; (i) Social honeypot dataset [2]. This dataset comprises a long-term study of social honeypots through 60 honeypots on Twitter that resulted in collecting 36,000 candidate content polluters. The labelled dataset is trained and tested over six popular machine learning algorithms. The performance of these algorithms are compared and their effectiveness is measured via a number of robust evaluation metrics. Further, the proposed approach is benchmarked against state-of-the-art social spam and anomalous detection techniques. These experiments verify the effectiveness and utility of the proposed approach and embedded modules.

The key contributions of this paper are summarised as follows:

- An effective multi-topic social spammer detection model for microblogging is proposed.
- The proposed model addresses the deficiencies in the existing approaches by conducting a fine-grained analysis to users data and metadata using semantic analysis and sentiment analysis., thereby extracting various topic-dependent and topic-independent features.
- The proposed model introduces a topic mapping scheme between two well-known topic discovery approaches.
- A comprehensive experimental analysis is conducted which verifies the utility of our model to detect social spammers in microblogging.

The remaining of this paper is organised as follows: the following section lays the background on various approaches used to detect spam in social networks and provides an evaluation of the existing approaches. Section 3 presents and discusses the methodology and embedded modules. The experiments conducted in this study are explained in Section 4. Section 5 discusses the key insights of this paper before we conclude it with certain remarks on future research directions.

## 2. Related Works

This section reviews important research on social spam detection. First, we discuss the notion of social spam followed by a report on various types of social spam. Then, recent solutions provided to tackle the social spam issue are presented and discussed.

### 2.1 Social spam – an overview

The Internet has witnessed a paradigm shift since the emergence of the second generation of the Web (i.e., Web 2.0, a.k.a. Social Web). This has transformed the Web experience enabling people to interact and exchange ideas, thoughts, and beliefs leveraging the free and easy access to such virtual environments. However, the lack of a 'gatekeeper' in these technological means has established a new form of spam, commonly referred to as social spam. The notion of social spam refers to any form of undesirable information (including textual content, images, videos, URLs, follow/friend requests, etc.) which spreads through a network(s) of social media in a process referred to as *spamming* [20]. Social spam differs from a typical spam email or SMS spam in which the former manifests in multiple forms and modi operandi, and the latter can be mainly conveyed in one form (i.e. emails [21-23] or SMSs [24]). The following are examples of social spam:

- *Social anomalies*: Despite the lack of consensus to provide a unified definition to the notion of an anomaly in OSNs, the term can be referred to as a deviation resulting from unexpected, illegal and irregular behaviour of users that is prevalent in these societies [25]. Producers of bad quality social data, such as anomalies, provide their content with anonymity and impunity.
- *Malicious links*: Links that cause harm to individuals or computers. They are embedded in the spear-phishing tweets and distributed through the networks [26].
- *Bulk messaging(mass direct messaging or spam-bombs)*: A chunk of messages of the same content that are proliferated to a group of users in a relatively short period of time [27].

- *Fake profiles (or Sybils/Socialbots)*: Social mock accounts that are created mainly to gain visibility or influence on social media. Social bots concoct various accounts of similar identities [28]. Twitter does not tolerate fake and suspicious accounts. They suspended more than 70 million accounts in 2018 [29]. Yet, it is expected that social bots will continue to propagate and social media manipulation remains undetected [30].
- *Fake reviews (opinion spamming)*: Reviews are written to promote or to discredit products, services or businesses, thus do not reflect a genuine user experience [31].

## 2.2 Approaches for social spam detection

The rapid increase in utilising OSNs along with the lack of the gatekeeper have simulated spammers to inject unsolicited contents into these platforms [32-37]. This has led the research community to pay more efforts to stop the spreading of spam and spammers. These endeavours are commonly divided into two main directions; (i) Machine Learning approaches: - The set of techniques that incorporate machine learning and artificial intelligence to detect and discover social spam; (ii) Graph-based approaches:- The techniques which rely on the social network structure and graph properties for spam detection.

### 2.2.1 Machine learning approaches –un/semi-supervised

Machine Learning algorithms have proven the ability to modernize technology enabling companies to design solutions to sophisticated problems as well as to make informed and better decisions[38-44]. The applications of machine learning span various domains such as healthcare [45], industry [46], recommender systems [47], and NLP systems [48]. The mechanisms of detecting spammers in OSNs incorporating machine learning are commonly determined based on the nature of training detection models [49, 50]. Supervised detection methods extract a collection of features that can be used to train the machine learning classification model [51]. These features are mainly inferred from the analysis of users' metadata and user's contents. For example, Clark et al. [52] incorporated certain features such as time between tweets, #followers, etc. to detect fake and robotic accounts. With a robust textual classification scheme, that uses natural language structure, authors were able to obtain promising results on discovering accounts that send automatic messages. Various supervised learning approaches that are built on labelled datasets and using classification algorithms such as Support Vector Machine, Logistic Regression, Decision Tree, and Naïve Bayes have proven effectiveness in spam detection [53].

The availability of publicly labelled datasets is very infrequent; thus, unsupervised learning methodologies provide an alternative as they do not require labelled datasets. For example, Zhang et al. [54] applied the Twice-clustering approach on product reviews collected from 360buy.com, and the authors obtained 66% accuracy for detecting spam reviews. Another thread of efforts incorporated a k-means clustering algorithm for spam detection. These efforts reported good results on spam detection. For example, Liu [55] and Wu [56] reported 71% and 72% accuracy respectively by employing the k-means approach. Unsupervised learning is also used in real-time spam tweet detection using collective-based tweets analysis [57], and also used in spam reviews detection [58] and bot detection [59].

The semi-supervised learning provides a trade-off between the aforementioned machine learning techniques; thus it can work with a few labelled observations along with more unlabelled ones. Amongst various techniques reported in the literature, generic or wrapper and non-generic are mainly key types of semi-supervised learning. For example, using a semi-supervised generative active learning approach to automatically generate semantically similar texts for spam content detection [60]. Another attempt was undertaken by [61] in which authors used semi-supervised learning over a partially labelled dataset for Twitter spam drift problem. Semi-supervised learning was further employed in [62-64]

### 2.2.2 Graph-based approaches

Social spam can also be detected using attributes and features captured from graph nodes of the social network. These features can be distilled from two categories; social features (i.e. account-based features) and structural features of the social graph [56]. Various studies were proposed in this direction. For example, Gupta et al. [65] developed a model to detect Twitter spammers who spread social spam by accessing phone numbers of Twitter users to deliver annoying advertisements of products and services. The authors proposed a Hierarchical Meta-Path Score (HMPS) metric to measure the similarity between two nodes in the network. Then, they framed a Twitter dataset as a heterogeneous network by leveraging diverse interconnections between different categories of nodes embedded in the dataset. Al-Thelaya et al. [66] proposed representation models for social interaction's graph-based datasets. The models were designed to detect social spam based on graph-based analysis and sequential processing of user interactions. Integrating both social content and network structure are present in the literature; Noekhah et al. [67] proposed MGSD (Multi-iterative Graph-based opinion Spam Detection) model to detect 'spamicity' effects of internal and external entities employing topic independent features Hybrid models were also demonstrated in [68-70].

2.2.3 Evaluation of current approaches

**Lack of incorporating semantic analysis:** The difficulty of acquiring an accurate understanding of the contextual meaning of social textual content affects a further conducted analysis. The existing social spam detection approaches, into their textual analysis, adopt a probabilistic generative model, namely the Latent Dirichlet Allocation (LDA) [11] and its variations such as Labeled LDA, PhraseLDA, etc. In spite of the popularity of these models to infer a predefined set of topics, they suffer from the following drawbacks; (i) they fail to capture high-level topics/domains; (ii) they are unable to consider and integrate the semantic relationships between terms in the text; and (iii) they are inadequate to extract topics from short textual contents such as tweets as well as to extract semantically meaningful data from discrete textual content [12, 13]. On the other hand, this study incorporates semantic analysis and semantic web techniques which enable the elicitation of meaningful information from social data, thereby enhancing its textual content to offer semantics and linking each message to a specific domain. IBM Watson NLU, for example, provides a list of taxonomies (categories) that are further subdivided into finer-grained subcategories based on ontologies and linked open data. Employing semantic analysis techniques have proven effective to address the brevity problem of textual messages such as tweets and to mitigate problems pertaining to features of linguistics such as polysemy, homonymy, and contronymy [37].

**Lack of topic-specific spam detection models:** Despite the ongoing endeavours to tackle the pressing problem of social spam and its widespread negative implications, the current approaches are inadequate to address multi-topic social spamming behaviours. The current social spam detection approaches are generic and passively extract features based merely on users' data and metadata. This study, on the other hand, proposes a novel and effective approach to detect social spammers incorporating a number of attributes to measure topic-dependent and topic-independent users' behaviours in OSNs. In particular, our approach proposes a novel topic distinguishing mechanism based on $tf.idf$ heuristic aspect of IR. This heuristic aspect is incorporated into our model to extract features from the users' contents based on their topics of interest. The extracted features from user's data are classified into two different groups (topic-dependent and topic-independent), thereby providing fine-grained topic-specific user behavioural analysis. The next section elaborates on the proposed methodology.

# 3. Methodology

Figure 1 illustrates the proposed framework and the embodied modules. As depicted in Figure 1, the system architecture comprises of three main phases which are detailed in the next sections, namely; (1) Tweets Acquisition and pre-processing; (2) Feature extraction and selection; and (3) Machine learning model building and tuning.

## 3.1 Tweets acquisition and pre-processing

This section presents the steps followed to collect and pre-process datasets. First, we provide an overview of the two Twitter sets used in this study, and then we discuss the undertaken steps for data preparation and integration.

### 3.1.1 Dataset source

This study focuses on social data generated from Twitter™. Twitter analytics is an evolving research field falling under the category of data mining, social big data analysis, and machine learning. Twitter has been chosen in this paper due to the following reasons: (i) Twitter platform has been studied broadly in the research communities [71], leveraging the vast volume of content (6,000 tweets/seconds) [72]; (ii) It facilitates retrieving public tweets through providing APIs; (iii) the Twitter messages' "max 140 characters" feature enables data analysis and prototype implementation for a proof of concept purpose.

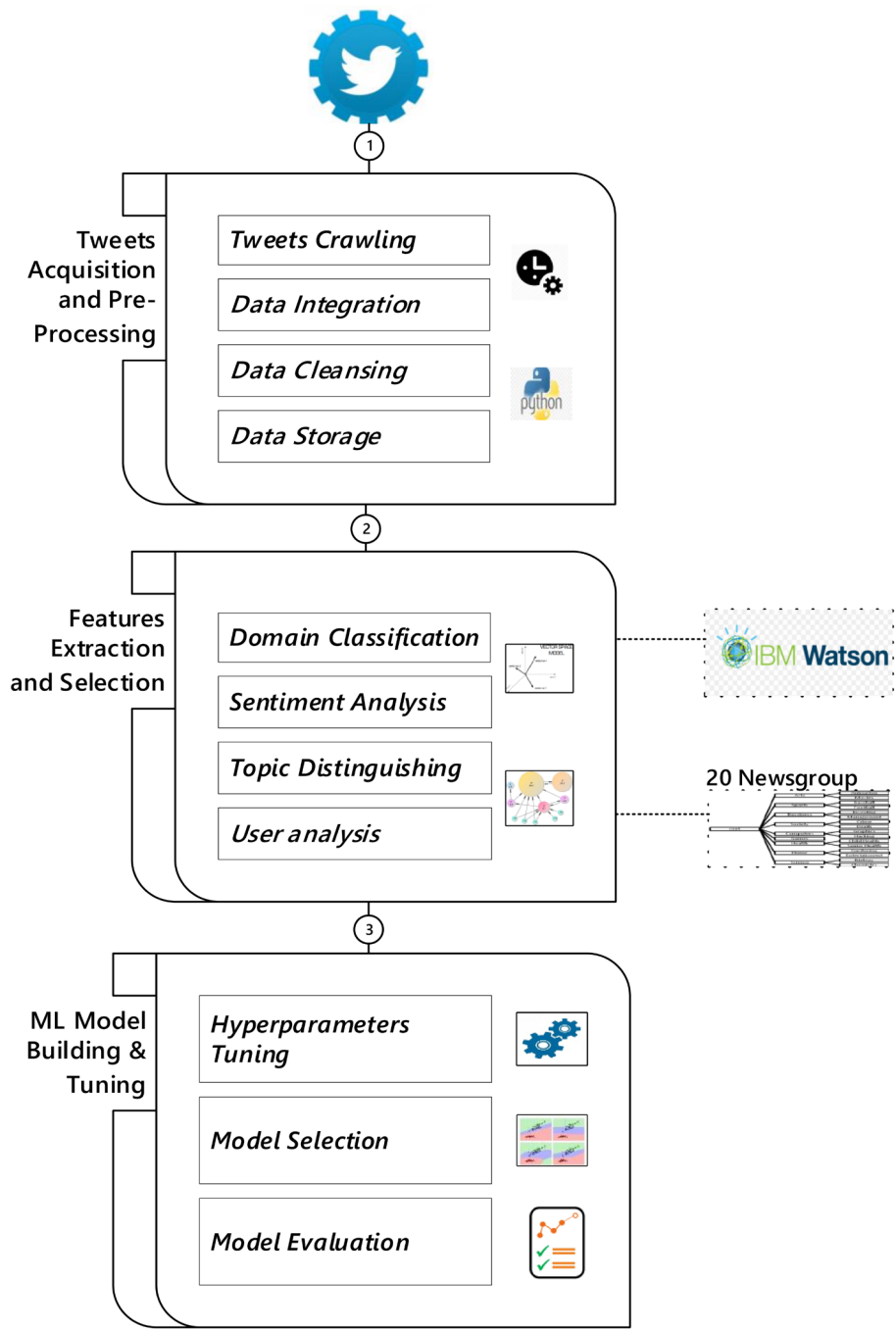

*Figure 1: System Architecture – prepared by the authors*

Twitter data access mechanisms have been harnessed in this study for data collection purposes. Users' information and their tweets and all related metadata were crawled using *TwitterAPI* [73]. A PHP script was implemented to crawl users' content and their metadata using the *User_timeline* API method. This API allows access and retrieve the collection of tweets posted by a certain user_id associated with each API request. This approach is used rather than a keyword search API due to the reasons as follows. Keyword-based search API has certain limitations listed in [71], i.e. Twitter index provides only tweets posted within 6-9 days thus it is hard to acquire a historical Twitter dataset before this period. Further, Search API retrieves results based on the relevance to the query caused by uncompleted results. This implies missing tweets and users in the search results. Using the user's timeline approach, on the other hand, retrieves up to 3,200 of the recent users' tweets. Last but not least, the purpose of this paper is to measure the users' credibility hence user-driven tweets collection is the suitable approach. Further, *acTwitterConversation*[1] API was used to retrieve all public conversations related to the tweets being fetched using Twitter API.

Data acquisition is carried out using a PHP script triggered by running a cron job that selects a new user_id and starts collecting historical user information, tweets, replies and the related metadata. The list of Twitterers' user_ids used in the data acquisition phase is extracted from two key selected datasets; (1) **Topically anomalous dataset**: A Twitter graph dataset crawled by Akcora et al. [19]. This graph is chosen since it includes the list of users who had less than 5,000 friends. This threshold was established by Akcora et al. [19] to discover bots, spammers and robot accounts; (2) **Social honeypot dataset:** This dataset comprises a long-term study of social honeypots through 60 honeypots on Twitter that resulted in the collecting of 36,000 candidate content polluters [2]. It is worth indicating that we have obtained only the list of user_ids from the aforementioned datasets. But the tweets and all the related contents and metadata were collected by our script, thereby studying the recent social behavior of those users as will be discussed in the experiments section. The collection of tweets was carried out using supercomputing facilities provided and supported by Pawsey Supercomputing Centre[2].

3.1.2 Tweets pre-processing and storage
Features of Big data should be tackled when handling large-scale dataset such as social data. To ensure the veracity of Big data in the context of this study, accuracy, correctness and credibility of data should be established. Although the origin of data and storage is critical to ensuring the veracity of Big data, the trustworthiness of the source does not guarantee data correctness and consistency. Data cleansing and integration should also be used to guarantee the veracity of data. Further improvements to the collected data quality will be discussed later in the analysis phase. The raw extracted tweets were subjected to a pre-processing phase to address the data veracity regarding data correctness. This phase includes the following steps:

**Data integration and temporary storage:** Tweets are collected from the designated APIs in JSON format. Further, the tweets' replies collected from AcTwitterConversation API is obtained in arrays. To maximize the reuse of data and facilitate data analysis, data integration aims to address the format dissimilarity by blending data from different resources. In particular, the raw tweets (JSON) and tweets' replies (ARRAY) are extracted, reformatted, and unified to fit a predesigned relational database ( i.e. MySQL) that is used as a temporal storage for data for the following analysis phase.

**Data cleansing:** Data at this stage may include various errors, meaningless and irrelevant data, redundant data, etc. Thus, data is cleansed to detect and remove corrupt, incomplete and noisy data, thus ensuring data consistency. For example, duplicate contents are removed. This comprises tweets, tweets' replies or any other metadata. Further, URLs to photos and videos are eliminated as those do not contain textual contents that can be extracted and examined in the proposed model. However, future research aims to provide a multimodal approach to tackle this issue.

---
[1] https://github.com/farmisen/acTwitterConversation
[2] https://pawsey.org.au/

## 3.2 Topic classification & sentiment analysis

In this module, we aim to segment users based on their topics of interest and to carry out opinion mining on the tweets' replies. For this purpose, we employ IBM Watson – NLU to be utilised for topic inference and sentiment analysis. Also, we develop a topic classification model using the 20 Newsgroups dataset.

**IBM Watson – Natural Language Understanding:** IBM Watson NLU is a cloud-based service which is used to extract metadata from textual content such as entities, taxonomies/categories/high-level topics, sentiments, and other NLP components. IBM Watson analyses the given text or URL and categorises the content of the text or webpage according to various topics/categories/taxonomies with the corresponding *scores* values. *Scores* are calculated using IBM Watson, range from "0" to "1", and report the accuracy of an assigned taxonomy to the analysed text or webpage. IBM Watson is used further to determine the overall positive or negative sentiment of a given text (reply). Table 1 demonstrates an example of incorporating this API to extract taxonomies and the sentiment of a given tweet. As illustrated in Table 1, the content of the tweet is analysed by IBM Watson using two key components: Categories Inference and Sentiment Analysis. The scores are given for each component to represent the adequacy of the retrieved taxonomy and sentiment to the provided tweets. The taxonomy inference module is used in this research in the topic discovery, while opinion mining is used to detect the sentiments of tweets' replies [74, 75].

*Table 1: An example of incorporating IBM Watson for taxonomies inference and sentiment analysis*

| Tweet | Categories Inference | | Sentiment Analysis | |
|---|---|---|---|---|
| "Even as we focus on fighting COVID-19, it's important to recognize that there's another pandemic raging right now—one that's decades in the making and unique to the United States. We need to treat gun violence with the same urgency and resolve." [3] | **Category** | **Score** | **Sentiment** | **Score** |
| | /health and fitness/disease/cold and flu | 0.909 | positive | 0.53 |
| | /health and fitness/disease/epidemic | 0.77 | | |
| | /society/unrest and war | 0.65 | | |

A tweet is commonly comprised of two key elements, namely the text and the embedded URL. URLs are attached to tweets to convey further information on the topic discussed in the tweets, especially that a tweet is limited in length. URLs direct users to another website, or an image, video, etc. Twitter automatically scrutinises URLs against a set of harmful and malicious sites, then those URLs are shortened to http://t.co links to help users share lengthy URLs into their short tweets. This feature is unfortunately abused by spammers through hijacking trends or embedding unsolicited mentions and attaching clickjacked URLs. This tricks twitterers into carrying out undesired actions by clicking on such concealed links. Therefore, as an important credibility measure, it is crucial to examine embedded URLs, detect topics of the URLs' web pages, and validate that topics match with the textual contents of tweets handling those URLs.

IBM Watson is utilised further in this study to stem the sentiment of a certain reply whether it is positive, natural, or negative with the corresponding sentiment score. Consequently, all of a tweet's set of replies are collected and the sentiments of these replies are combined in the analysis to enhance the credibility.

**Topic discovery using the 20 Newsgroup:** The 20 Newsgroups data set is a collection of around 20,000 collected news documents which are grouped into 20 high-level categories (newsgroups) [18]. Each document in this corpus is labelled with one of the twenty categories. This dataset is a popular dataset and is commonly used to conduct textual clustering tasks. The documents embedded in this set are distributed almost evenly amongst the twenty domains, and each document is written in a form of an email. This dataset is further used to provide consolidate the topic discovery process. To use the dataset in the intended task, the textual contents of the dataset were vectorized into numerical vectors to be used for conducting the predictive task. In this regard, a Bag-of-Words technique, namely TF-IDF is used to extract feature vectors from each of the textual documents. The read-for-prediction dataset is then used to feed the incorporated classification modules. Further details will be discussed in the experimental results section.

---

[3] This is a tweet by Barak Obama on 24/02/2021

**Topic mapping scheme:** The aforementioned topic discovery and classification approaches provide different high-level domains, yet a unified model is required to facilitate features extraction to be used for the credibility module. Therefore, we propose a mapping scheme between all domains of the two approaches. In particular, we manually and carefully match domains in two approaches if they are semantically interrelated. Figure 2 shows the proposed mapping between all of the two high-level categories.

This mapping establishes a unified model which will be used in the topic analysis of the users' textual content. Hence, textual contents of the tweets will be analysed using two approaches and only tweets that meet the following criteria will be selected: A tweet will be selected if the extracted topic in each module corresponds to the proposed mapping of Figure 2. Thus, tweets that infer different domains do not comply with the designated mapping scheme and will be neglected and not be incorporated in a further conducted analysis.

The aim of incorporating two different mechanisms for topic discovery is twofold: (i) to consolidate the inferred topic by establishing an agreement between more than one topic discovery model; and (ii) to validate and verify the need for more than one topic discovery model in similar tasks. For example, poor correlation in the number of matched topics extracted by analysing the same tweets poses a question on the utility of incorporating a sole model for a topic discovery task. This aspect is further elaborated in the experimental results section.

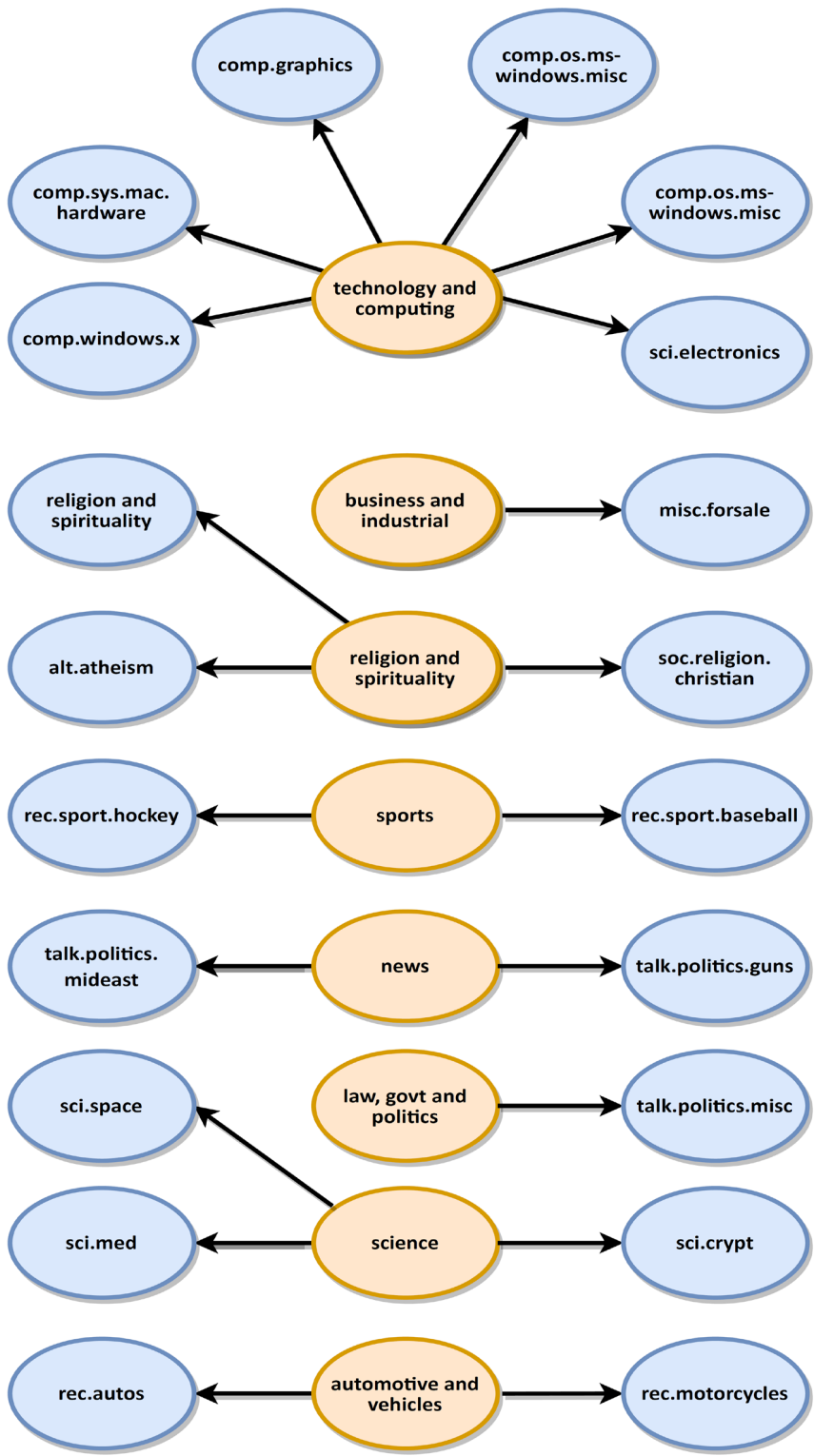

*Figure 2: Topic mapping between 20-Newsgroup and IBM Watson NLU-Categories*

## 3.3 Features extraction and selection

Understanding users' behaviour in the context of social networks is important to detect and identify various categories of users in these virtual environments. Spammers exhibit different behaviour than normal and legitimate users. This can be seen in different activities such as posting activities, following rate, quantity and quality of shared content and URLs, etc. In this study, we examine spammers behaviour both at the content level and at the user level. At the content level, we perform fine-grained analysis to users' content to examine and count all topics obtained from their textual content, hashtags and embedded URLs. Further, we listen to the voice of users' followers by conducting sentiment analysis to their followers' replies. At the user level, we examined the following to follower relationship incorporating the account's age. Also, other fine-grained metrics extracted from user metadata is examined and utilised in the endeavour of evaluating users' credibility. Features and mechanisms used to extract them are discussed in the following sections.

### 3.3.1 Topic distinguishing mechanism

The analysis of a user's content to discover the user's main topics of interest is an essential start to the process of measuring the user's credibility. In social networks, a user conveys to others his/her interest in a certain topic(s) by continuously posting tweets and attaching URLs with those tweets to discuss that particular topic(s). This category of users should be given a higher weight than the other category of users whose tweets discuss a various number of topics, thus their topics of interest can be difficult to be identified. To quantify this weight, the theoretical notion of TF-IDF has been used to distinguish users' topics of interest.

TF-IDF is considered as a fundamental component embodied onto several Information Retrieval models, namely the vector space model (VSM) [76]. "The intuition was that a query term which occurs in many documents is not a good discriminator" [15]. This indicates that a word that occurs in several documents reduces its weight in the document and the corpus as this word does not help with the documents retrieval process as it does not reveal the document(s) of interest to the user [16]. TF-IDF is commonly used to quantify the significance of a word to a certain document in a collection of documents. It includes standard concepts which formulate its structure; Term Frequency (TF): is used to compute the number of times that a word appears in a certain document. TF, therefore expresses the importance of a word in the document; Document Frequency (DF): evaluates the significance of a word to a document in the overall corpus [77], and Inverse Document Frequency (IDF): is used as a distinguishing measure for a word in the text corpus to infer the term's importance in a certain document(s) [14]. Therefore, TF_IDF integrates the definitions of the significance of each index word in the document and the importance of the index term in the text collection to produce an aggregate weight for each word in each document. It assigns to a word $w$ a weight in document $d$ that is: (i) greatest when $w$ appears several times within a few numbers of documents; (ii) smaller when the word $w$ appears less times in a the document $d$, or appears in several documents; and (iii) smallest when the word $w$ appears in all documents of the corpus.

In the context of this research, this heuristic notion is assimilated into our framework to measure the credibility of users. Thus, it is argued that a user who posts on all topics has low credibility in general. This argument can be justified based on the following facts: (i) No one person is an expert in all topics [8]; (ii) A user who posts in all topics does not state or convey to other users which topic(s) s/he is interested in. The topic of interest can be conveyed by posting a wide range of content in that particular topic; (iii) There is a possibility that this user is a social spammer due to the behaviour of spammers posting tweets on several topics [9]. This could end up posting tweets on all topics which does not demonstrate a legitimate user's behaviour.

### 3.3.2 Incorporated features

A set of various features is used and extracted from both users' data (users' tweets) and their metadata (miscellaneous data collected from Twitter APIs). These features can be categorised into topic-dependent and topic-independent features.

**Definition (1).** Topic dependent features are the set of fine-grained features in which values are computed considering each topic of interest captured from the user's content.

**Definition (2).** Topic independent features are the set of coarse-grained features in which values are computed with no consideration is taken regarding the user's topic of interest.

Table 2 shows the list of both topic-dependent and topic-independent features incorporated in this study along with their description.

Table 2: Incorporated Features (domain-dependent and domain-independent)

| No. | Feature | Description | Topic Dependent |
|---|---|---|---|
| $x_1$ | **#Words** | The count of words in each topic captured from the user's tweets. | √ |
| $x_2$ | **#Unique words** | The count of non-redundant words in each topic obtained from the user's tweets. | √ |
| $x_3$ | **#URLs** | The total count of URLs posted in user's tweets of each topic. | √ |
| $x_4$ | **#Unique URLs** | The total count of non-redundant URLs posted in user's tweets of each topic. | √ |
| $x_5$ | **#Unique URLs' hosts** | The total count of non-redundant URLs posted in user's tweets of each topic. | √ |
| $x_6$ | **Topic frequency** | The total number of topics the user has tweeted about. | √ |
| $x_7$ | **Inverse topic frequency** | Distinguishes users among the list of their topics of interest. It is computed as: $$IDF_u = log(\frac{n}{DF_u})$$ Where $n$ is the number of topics that a user discussed, $DF_u$ is the topic frequency. | √ |
| $x_8$ | **#User's retweets** | The total number of retweets for user' content in each topic. | √ |
| $x_9$ | **#User's likes** | The total number of likes count for the users' content in each topic. | √ |
| $x_{10}$ | **#User's replies** | Embody the number of replies to the users' content in each topic. | √ |
| $x_{11}$ | **Sum of user positive sentiment replies** | Refers to the total sum of the positive scores of all replies to each user in each topic. Positive scores are achieved from IBM Watson NLU (sentiment analysis) with values greater than "0" and less than or equal to "1". The higher the positive score, the greater is the positive attitude the repliers have to the users' content. | √ |
| $x_{12}$ | **Sum of user negative sentiment replies** | Represents the sum of the negative scores of all replies to a user u in each topic. Negative scores are those values greater than or equal to "-1" and less than "0". The lower the negative score, the greater is the negative attitude the repliers have to the users' content. | √ |
| $x_{13}$ | **Users' followers** | The total count of users' followers. | |
| $x_{14}$ | **User's friends** | The total count of the user's friends (followees) | |
| $x_{15}$ | **Followers-friends ratio** | User followers-friends ratio. It is computed as: $$\begin{cases} \frac{FOL_u - FRD_u}{FOL_u} \times \frac{Age_u}{100}, & if\ FOL_u - FRD_u > 0 \\ \frac{1}{FOL_u} \times \frac{Age_u}{100}, & if\ FOL_u - FRD_u \leq 0 \end{cases}$$ where $Age$ is the age of user profile in years. | |
| $x_{16}$ | **#Hashtags** | The total number of hashtags in the user's tweets. | |
| $x_{17}$ | **#DistinctHashtags** | The total count of the unique number of hashtags in users' tweet. | |
| $x_{18}$ | **Topic-specific Hashtags Ratio** | Indicates the % between the number of hashtags mentioned in the tweets to the total number of tweets for each user in each topic. | √ |

# 4. Experimental Results

## 4.1 Dataset selection and exploration

As discussed in the dataset source section, this study incorporates a set of users collected from two different sources, namely: the topically anomalous dataset [19] and the social honeypot dataset [2]. This experiment is undertaken on 4,000 users (2,000 from each source) amongst users with the highest number of tweets in each set. As depicted in the system architecture, Figure 1, all historical tweets of the selected users were collected and preprocessed as discussed in Section 3.1.2. Figure 3 shows the distribution of the collected tweets for the period between 2009 and 2020 which are related to the users obtained from the two datasets. In comparison with the continuous increase in the tweets count of the topically anomalous dataset, it is interesting to notice the decline in the posted tweets for users in the honeypot dataset. This can be interpreted as the latter dataset were intentionally containing polluters of social contents, thereby the number of social spammers of this dataset can be intuitively higher than the former dataset which contains users whose only suspicious feature is that the number of followers is less than 5,000. Therefore, we would expect this decline in posted tweets by users of the topically anomalous dataset as they can be detected by Twitter and thus banned from the Twitter community.

This study supports this endeavour and provides another approach to detect this category of users that can hide from Twitter's radar.

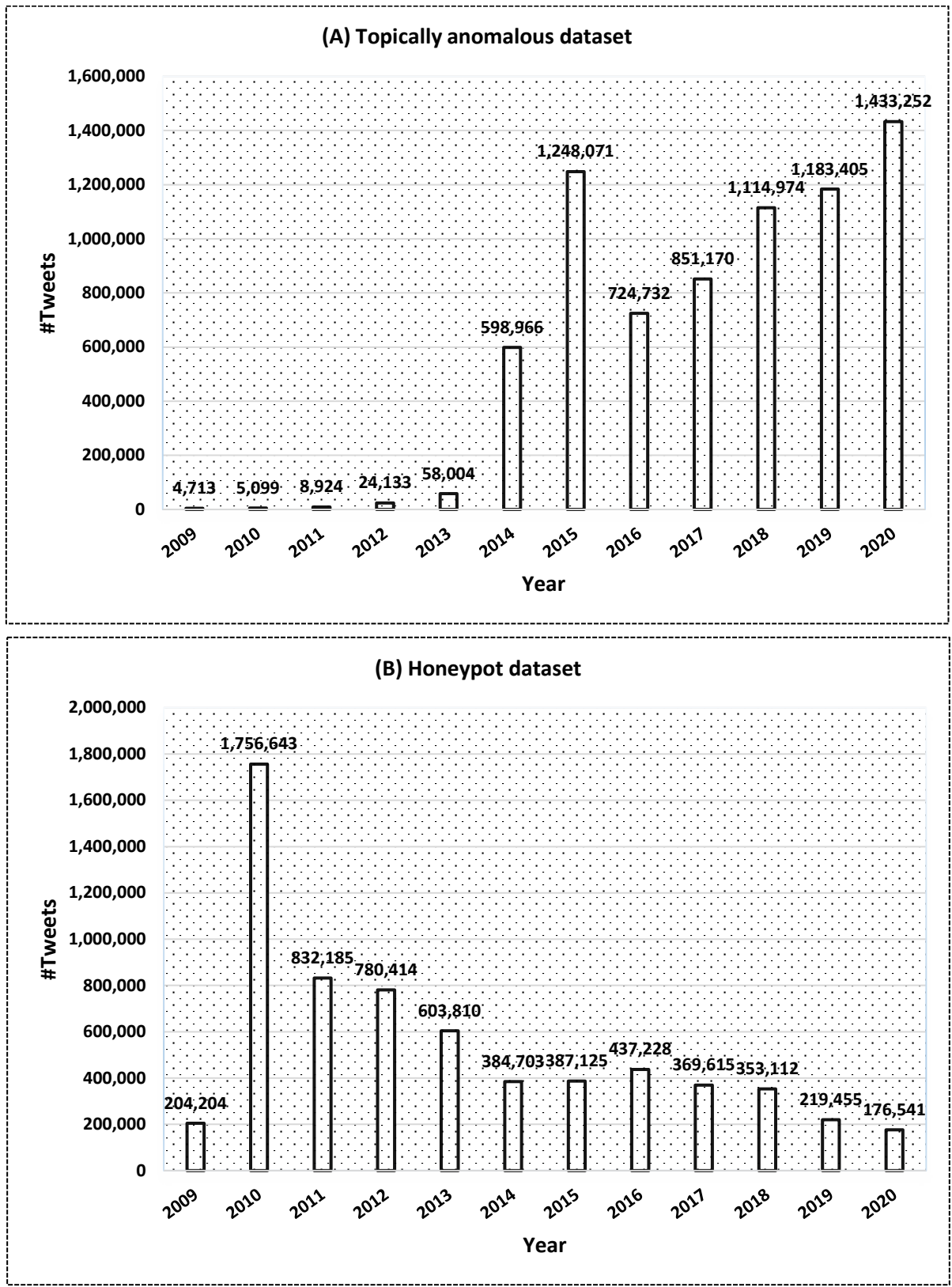

*Figure 3: Tweets distribution between 2009 and 2020 for the two selected datasets*

All tweets and users' metadata of both datasets are examined through written python scripts which were used to shorten all embedded URLs, extract URLs' hosts and all hashtags, and conduct all the experiments of this study. Table 3 shows a range of statistics on the users and their metadata collected for this study. This contains the collected tweets with their replies/conversations, categories/topics captured for those tweets, #URLs and

#unique host URLs embedded in the tweets. Also, both datasets contain the #words, #unique words, #hashtags, and #unique hashtags. All these figures are obtained from all historical tweets up to 2020.

*Table 3: Statistics of the collected datasets*

| Data Source | #users | #tweets | #tweets replies/ conversations | #URLs | #unique URLs | #unique host URLS | #words | #unique words | #hashtags | #unique hashtags |
|---|---|---|---|---|---|---|---|---|---|---|
| Topically anomalous dataset [19] | 2,000 | 6,270,257 | 114,481 | 1,238,395 | 307,855 | 40,133 | 83,432,897 | 19,178,212 | 1,904,985 | 403,272 |
| Honeypot dataset [2] | 2,000 | 5,463,761 | 752,297 | 1,451,511 | 457,489 | 147,342 | 75,763,023 | 24,052,466 | 2,061,820 | 865,734 |

## 4.2 Topics extraction

As discussed in Section 3, all historical tweets of all selected users were collected and examined through two topic discovery modules, namely IBM Watson NLU API as well as a developed classification model using 20-Newsgroup. IBM Watson NLU was used as an off-the-shelf API to infer the topic(s) of the examined pre-processed textual contents of the tweets and the websites of the embedded URLs. To consolidate the topics obtained by IBM Watson NLU API, we developed a topic classification model based on the 20-Newsgroup data set. In this task, the Python scikit-learn's implementation of textual classification using 20-Newsgroup[4] is used and enhanced. In particular, besides the Multinomial Naïve Based (Multinomial NB) classifier that is used by scikit-learn, two more classifiers are implemented, namely Logistic Regression (LR) and Stochastic Gradient Decent (SGD). Further, grid search for the optimal hyperparameters is carried out to find optimal and tuned parameters. The outcome of this experiment demonstrates that SGD outperforms other classifiers with an accuracy of 0.91 which is higher than the accuracy obtained by the scikit-learn's default implementation of Multinomial NB (≈ 0.77).

Figure 4 demonstrates the distribution of the tweets over each topic based on 20-Newgroup for two subsets: (A) Topically anomalous dataset, and (B) Social honeypot dataset, and Figure 5 illustrates the distribution of the tweets over topics captured using IBM Watson NLU for the same datasets. The next step is to capture the mapping between the two topic discovery approaches, thereby finding the tweets that match topics as illustrated in Figure 2 and discussed in section 3.2. Table 4 and Table 5 demonstrate the total number of tweets in each topic and corresponding topic using the incorporated domain discovery approaches on both datasets. Interestingly, these tables show a relatively poor correlation between domains captured using both topic discovery models. This divergence in capturing common topics in both models poses a question on the mechanism used by such well-known models to infer high-level topics. In future research, we will further examine this poor correlation by conduction a comprehensive study on similar high-level topic discovery models to verify the utility of such approaches in topic discovery designated tasks.

---

[4] https://scikit-learn.org/0.19/datasets/twenty_newsgroups.html

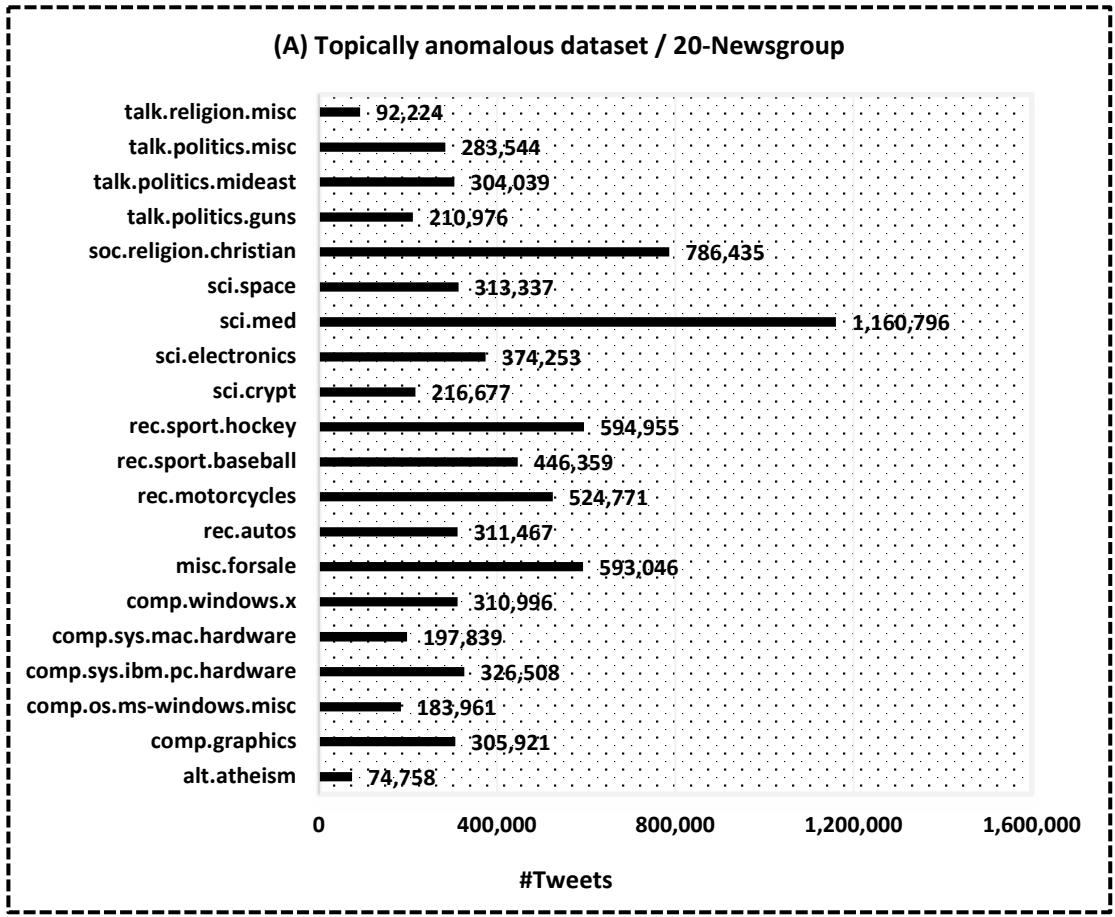

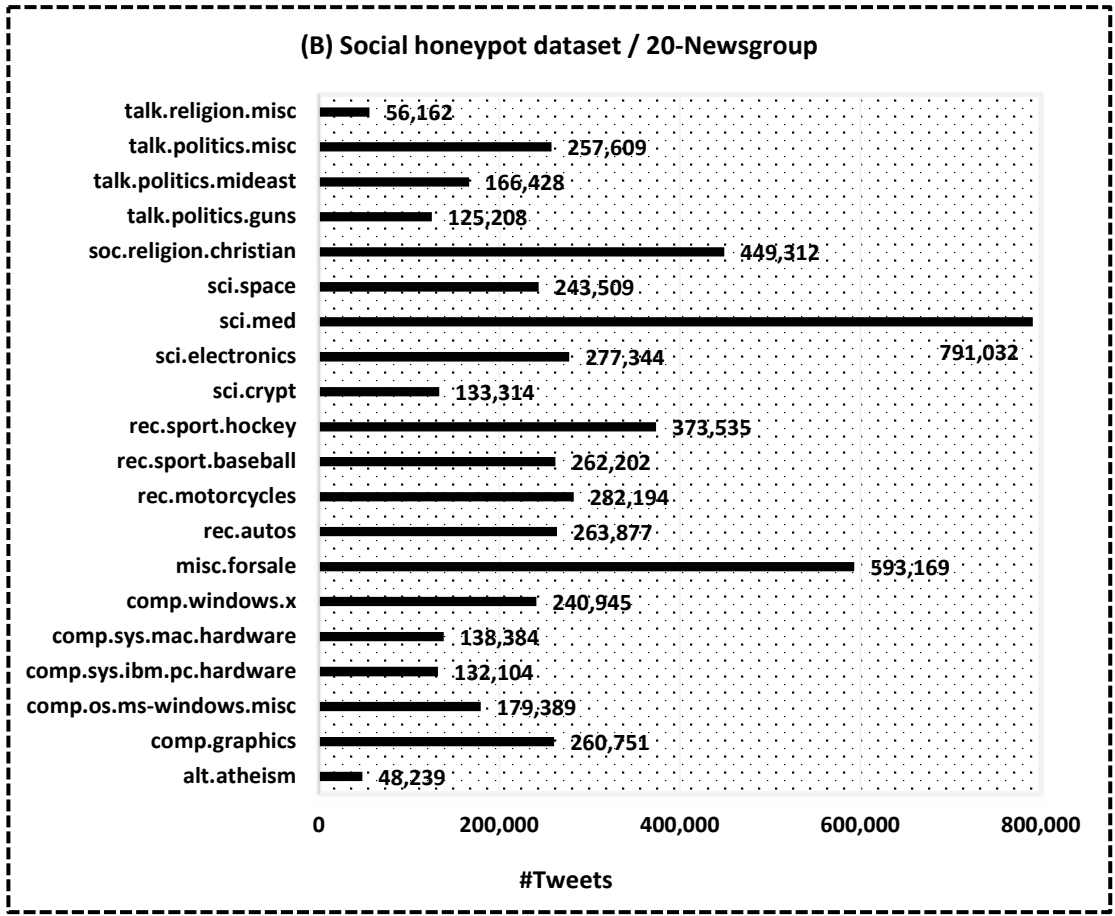

*Figure 4: The distribution of the tweets in each topic based on 20-Newgroup for two subsets: (A) Topically anomalous dataset, and (B) Social honeypot dataset.*

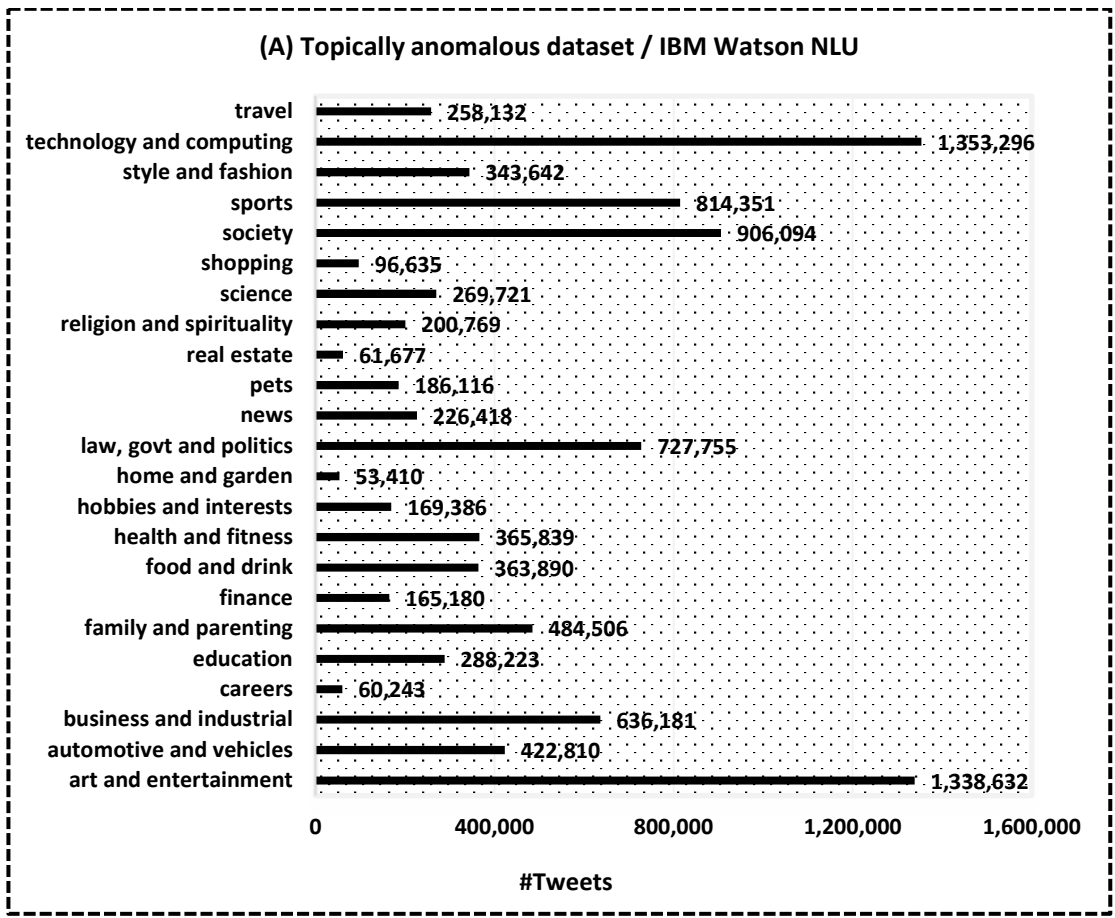
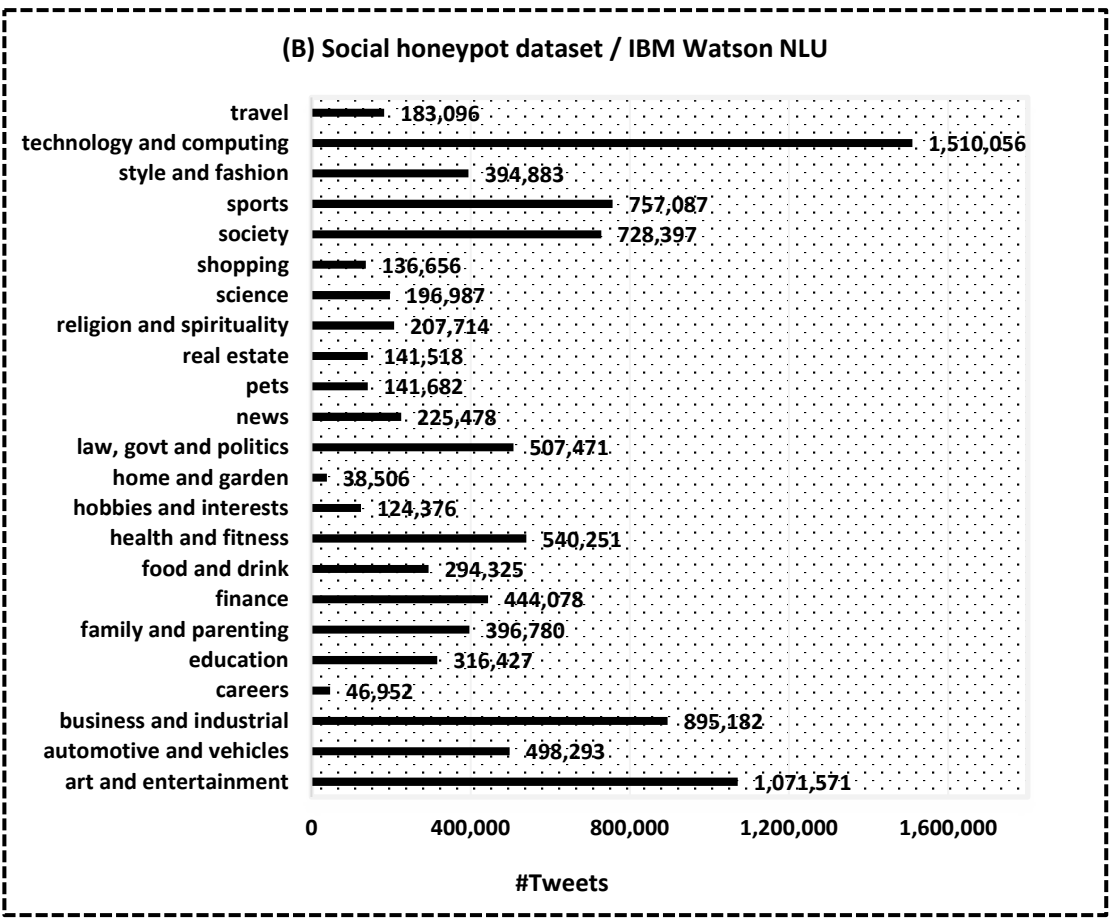

Figure 5: The distribution of the tweets in each topic based on IBM Watson NLU API for two subsets: (A) Topically anomalous dataset, and (B) Social honeypot dataset.

*Table 4: The tweets distribution over domains captured using two topic discovery approaches on the topically anomalous dataset.*

| IBM Watson NLU topics | #Tweets | Corresponding topcis(s) using 20-Newsgroup | #Tweets | #Matched Tweets | %Matching |
|---|---|---|---|---|---|
| technology and computing | 1,353,296 | comp.graphics | 305,921 | 473,627 | **35%** |
| | | comp.os.ms-windows.misc | 183,961 | | |
| | | comp.sys.ibm.pc.hardware | 326,508 | | |
| | | comp.sys.mac.hardware | 197,839 | | |
| | | comp.windows.x | 310,996 | | |
| | | sci.electronics | 374,253 | | |
| business and industrial | 636,181 | misc.forsale | 593,046 | 60,126 | **10%** |
| automotive and vehicles | 422,810 | rec.autos | 311,467 | 105,119 | **25%** |
| | | rec.motorcycles | 524,771 | | |
| sports | 814,351 | rec.sport.baseball | 446,359 | 257,998 | **32%** |
| | | rec.sport.hockey | 594,955 | | |
| science | 269,721 | sci.crypt | 216,677 | 64,203 | **24%** |
| | | sci.med | 1,160,796 | | |
| | | sci.space | 313,337 | | |
| news | 226,418 | talk.politics.guns | 210,976 | 18,595 | **8%** |
| | | talk.politics.mideast | 304,039 | | |
| law, govt and politics | 727,755 | talk.politics.misc | 283,544 | 72,693 | **26%** |
| religion and spirituality | 200,769 | talk.religion.misc | 92,224 | 75,760 | **38%** |
| | | alt.atheism | 74,758 | | |
| | | soc.religion.christian | 786,435 | | |

*Table 5: The tweets distribution over domains captured using two topic discovery approaches on the social honeypot dataset.*

| IBM Watson NLU topics | #Tweets | Corresponding topcis(s) using 20-Newsgroup | #Tweets | #Matched Tweets | %Matching |
|---|---|---|---|---|---|
| **technology and computing** | 1,510,056 | comp.graphics | 260,751 | 465,558 | **81%** |
| | | comp.os.ms-windows.misc | 179,389 | | |
| | | comp.sys.ibm.pc.hardware | 132,104 | | |
| | | comp.sys.mac.hardware | 138,384 | | |
| | | comp.windows.x | 240,945 | | |
| | | sci.electronics | 277,344 | | |
| **business and industrial** | 895,182 | misc.forsale | 593,169 | 85,194 | **14%** |
| **automotive and vehicles** | 498,293 | rec.autos | 263,877 | 137,915 | **28%** |
| | | rec.motorcycles | 282,194 | | |
| **sports** | 757,087 | rec.sport.baseball | 262,202 | 199,620 | **31%** |
| | | rec.sport.hockey | 373,535 | | |
| **science** | 196,987 | sci.crypt | 133,314 | 48,608 | **25%** |
| | | sci.med | 791,032 | | |
| | | sci.space | 243,509 | | |
| **news** | 225,478 | talk.politics.guns | 125,208 | 17,108 | **8%** |
| | | talk.politics.mideast | 166,428 | | |
| **law, govt and politics** | 507,471 | talk.politics.misc | 257,609 | 44,818 | **17%** |
| **religion and spirituality** | 207,714 | talk.religion.misc | 56,162 | 66,791 | **32%** |
| | | alt.atheism | 48,239 | | |
| | | soc.religion.christian | 449,312 | | |

The set of key features as depicted in Table 2 is extracted from users' data and metadata. The topic-dependent features are computed based on the 'Information Technology' topic. In future, we will carry out a cross-topic comparison between all features that are extracted based on all available topics. The Pearson correlation between incorporated features is computed as illustrated in Figure 6. The correlation matrix as depicted in Figure 6 is acceptable and most of the features are not highly positively correlated, thus no redundancy can be observed amongst features that might affect the employed models' performance. We incorporate lasso regularization [78] method to carry out automatic feature selection to enhance accuracy and interpretability.

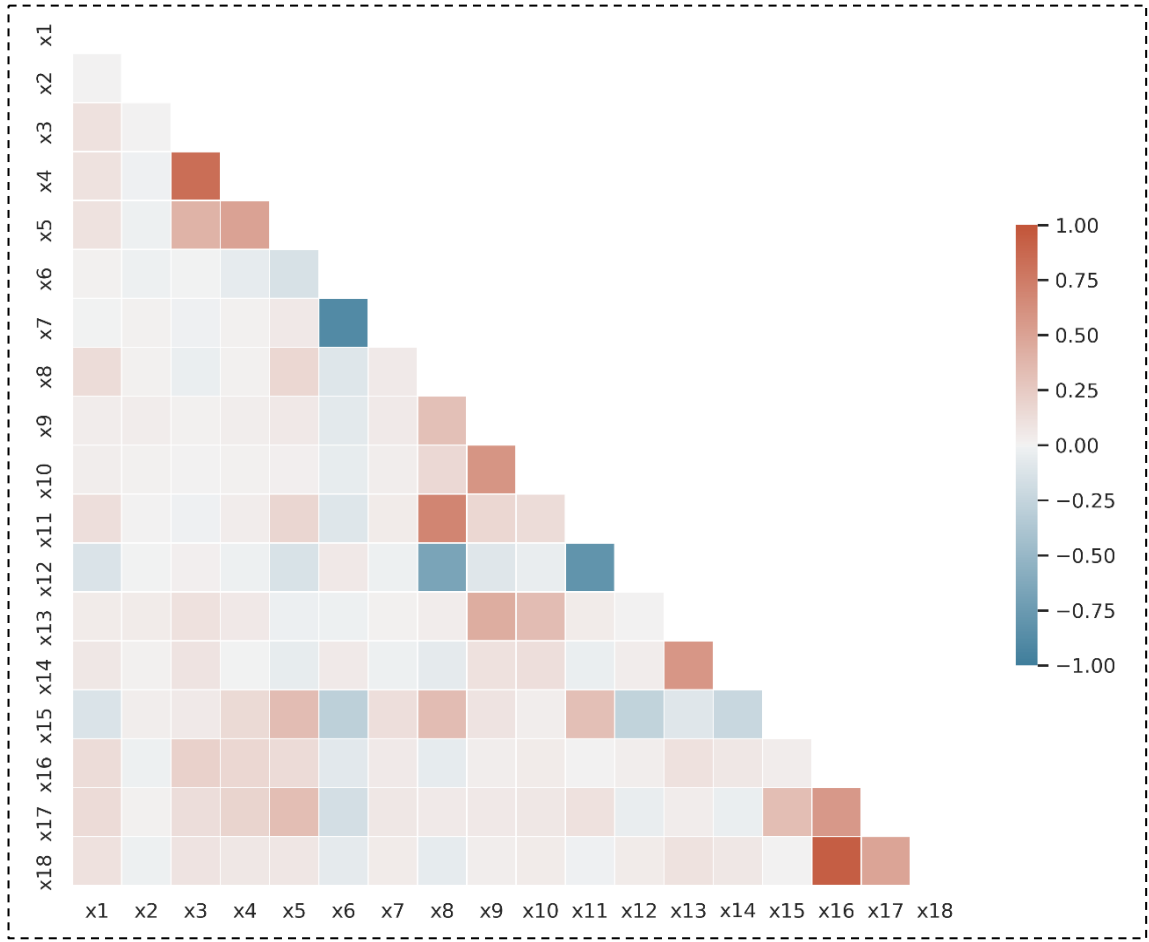

Figure 6: Correlation of incorporated features

The collected users were then manually and carefully examined to label and identify those who demonstrate spamming behaviour, thereby establishing a ground truth data set for measuring the performance of the proposed system based on the extracted features. Figure 7 shows the percentage of spammers to non-spammers users in the collected and labelled datasets. As depicted in Figure 7, only 37.75 per cent of the examined users are labelled as spammer users even though the selected datasets embody users who should be highly exhibiting spamming behaviour. This indicates the significance of conducting further analysis on published spam datasets.

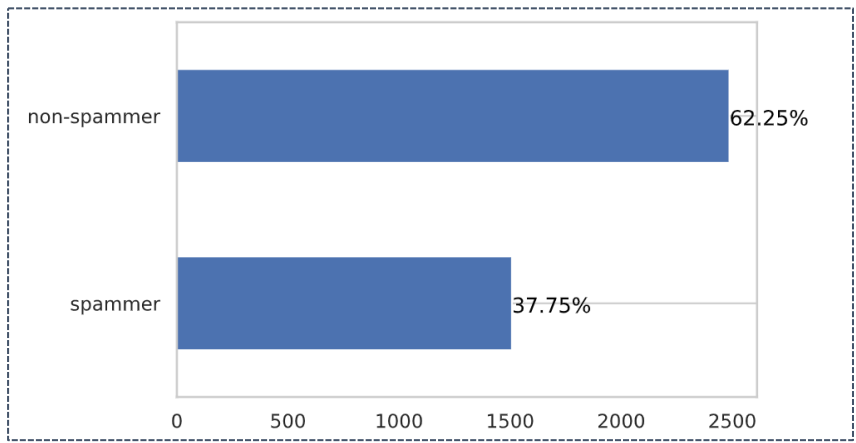

Figure 7: The percentage of spammers to non-spammers in the labelled dataset

The correlation depicted in Figure 8 demonstrates the importance of each feature and how each feature relates to the designated class label. As indicated in Figure 8 certain features play a significant role to infer those demonstrating spamming behaviour amongst the second category. For example, *Topic frequency ($x_6$)* shows that spammers tend to publish contents related to many topics of knowledge. Also, legitimate and spammers behaviours differ in terms of *#retweets ($x_8$)*, *#likes ($x_9$)*, and *#replies ($x_{10}$)*. These conventional features have been continuously proving their utility in detecting users with spamming behaviours. The last three boxplots of Figure 8 namely, *#followers ($x_{14}$) #friends ($x_{15}$), and followers-friend-ratio ($x_{16}$)* demonstrate the importance of incorporating age of the user's profile. In particular, the median of both *#followers ($x_{14}$)* and *#friends ($x_{15}$)* are close in values - despite the observed outliers of non-spammer users which can be understood due to the fact that this is a legitimate category of users in which we can find users with a high number of followers. Yet the convergence in the median of *#followers ($x_{14}$)* and *#friends ($x_{15}$)* for both spammers and non-spammers categories poses a question on the utility of models that mainly rely on these features. To tackle this issue, we proposed the *followers-friend-ratio* ($x_{16}$) which involves the age of the profile in computing its value as depicted in Table 2. This ratio credits the older profiles as they likely belong to legitimate users, and the correlation depicted in the last boxplot of Figure 8 justifies this claim.

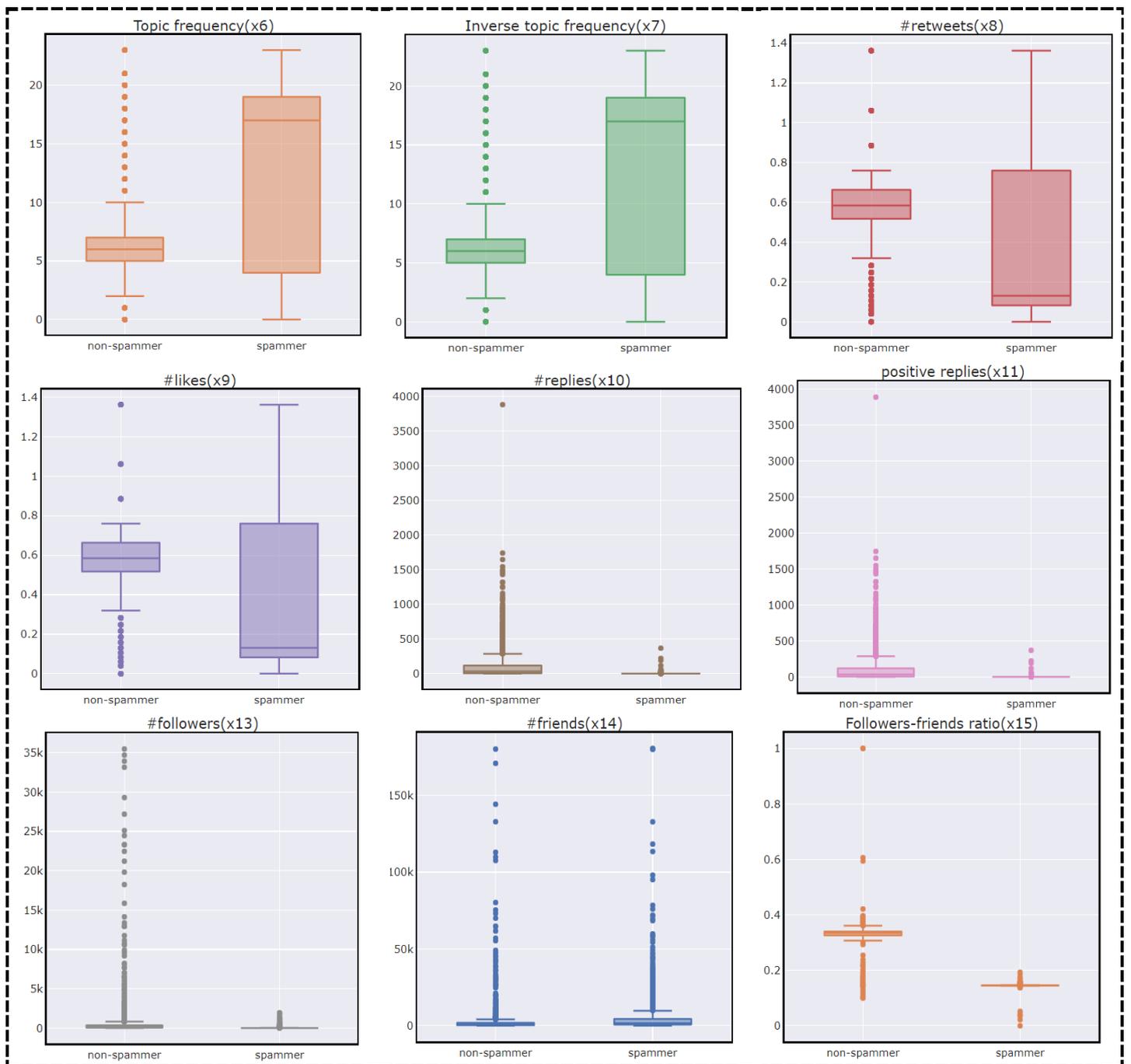

*Figure 8: Correlation between selected incorporated features and class label.*

To undertake further exploration, we also carry out bivariate distribution on '*topic frequency ($x_6$)*' feature to obtain a better understanding of the predictive power of topic frequency to the designated label. Figure 9 depicts dissimilar perspectives which conclude that the topic frequency feature is predictive. For example, the first chart of Figure 9 shows the densities of two categories in the dataset (spammers and non-spammers). These densities convey diverse distributions which imply pattern differences. It can be observed that most of the non-spammer users discuss an average number of topics which explains why the density of non-spammers is positioned in the middle of the chart. On the other hand, the spammers demonstrate an interest to post in few specific topics [79] or many topics [10], thus the density is positioned on the far-left and far-right sides of the chart. The subsamples in the middle chart (*bins*) establish the composition of various groups of topic frequencies. The dissimilarity in the proportion of spammers and non-spammers in each group validates again the significance of topic frequency as an important measure for prediction. Spammers tend to have a high topic frequency or low topic frequency which confirms the utility of incorporating this theoretically proven and intuitive aspect in this study. In particular, there is an inverse relationship between the number of topics the user is interested in and the user's credibility. This argument is justified based on the following facts: (i) there is no well-informed legitimate user who has the intellectual capacity to publish contents in all topics [8]; (ii) users who publish contents on various topics of interest do not convey to other users the particular topic that they are interested in. This is evident as topics of interest can be intuitively deduced from content of users who commonly post wide-ranging content in one or few topics; (iii) it is likely that this user is a spammer or anomalous; this illegitimate category of users tends to post tweets about numerous topics [9]. This could end up with tweets being posted on all topics which do not convey a legitimate user's behaviour. Therefore, distinguishing users in a set of topics is a significant aspect. The box plots appear in the third chart spot the behaviours of anomalies inferred from the dataset. These three charts prove the importance of topic frequency as an important variable in this study.

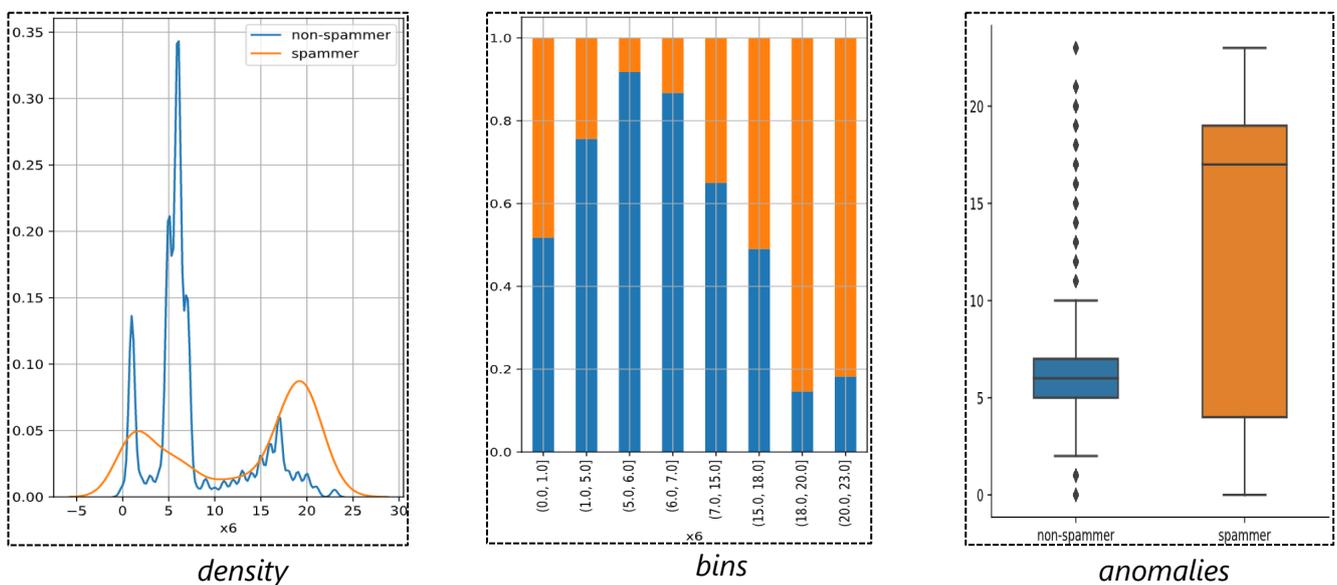

*density*     *bins*     *anomalies*

*Figure 9: Bivariate distribution on 'topic frequency ($x_6$)'*

### 4.3 Social spam system evaluation

#### 4.3.1 Evaluation metrics

The proposed system framework can be incorporated to classify whether or not a user is anomalous. As depicted in Figure 10, the proposed system framework can be used to determine (classify) if a user is a spammer or non-spammer. Four scenarios are illustrated by the classification as (i) a **true positive** ($tp$): is an actual spammer user where the model correctly predicts it as a spammer user; (ii) a **true negative** ($tn$) is a non- spammer (legitimate) user where the model correctly predicts it as a non- spammer user; (iii) a **false positive** ($fp$) is a non- spammer user where the model incorrectly predicts it as a spammer user; (iv) a **false negative** ($fn$) is an spammer user where the model incorrectly predicts it as a non-spammer user.

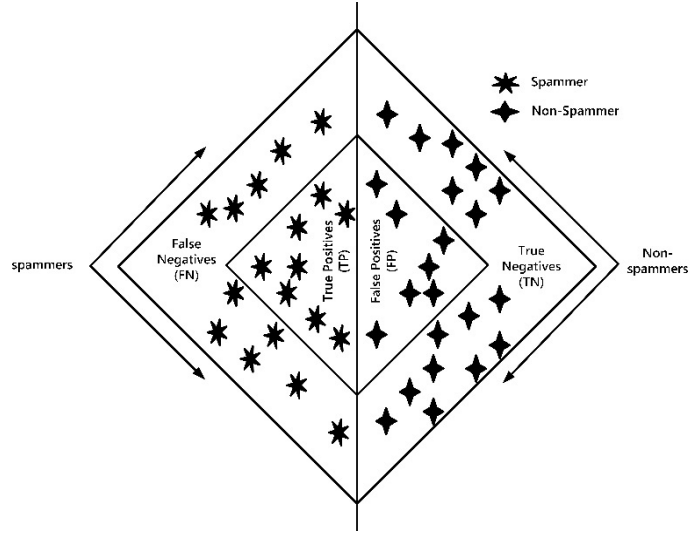

*Figure 10: Classification of spammers/non-spammers users*

The experiment of this study has been carried out incorporating various machine learning algorithms including Random Forest Classifier (RF), Multi-Layer Feed-Forward Neural Network (MLFFNN), Gradient Boosted Trees (GBT), Decision Tree (DT), Generalised Linear Model (GLM) and Naïve Bayes (NB). Although selecting a machine learning model for a certain classification task remains an ad-hoc process that requires further scrutiny, the selected models are well-proven learning algorithms and were selected due to their utility in various applications, especially in binary classification tasks [35, 80-82]. The performance of each of the implemented machine learning models in predicting multi-topic spammer users is measured based on theoretically proven key metrics, namely, *accuracy, precision, recall and F-measure*.

- **Accuracy:** it is commonly used for determining the rate of correct classification obtained by the incorporated algorithm. It is computed as:

$$accuracy = \frac{tp + tn}{fn + tp + fp + tn}$$

- **Precision:** indicates the proportion of correct anomalous predictions to those who are actually anomalous. Thus, it provides the rate of anomalous users who are classified correctly as anomalous. Precision can be calculated as:

$$precision = \frac{tp}{tp + fp}$$

- **Recall**: indicates the proportion of actual anomalous who were classified correctly as anomalous. Recall is computed as:

$$recall = \frac{tp}{tp + fn}$$

- **F-score:** conveys the trade-off between both Recall and Precision. Its formula is:

$$F - Score = \frac{precision \cdot recall}{precision + recall} = \frac{tp}{tp + \frac{1}{2}(fp + fn)}$$

Accuracy is considered the most intuitive performance metric as it is used in this study to infer the ratio of correctly predicted observations to the total observations (samples). Accuracy is a good performance measure;

thus a model of a high accuracy might indicate a well-performed model, yet this is true when the dataset is symmetric (class-balanced) and embodies an almost equal number of false positives and false negatives. The disparity between the number of spammers and non-spammers in our dataset requires incorporating Precision and Recall metrics. Precision indicates the ratio of the accurately predicted spammers to the total predicted spammers (accuracy of minority class predictions). This measure answers the question: Amongst all users labelled as spammers in the labelled dataset, how many users are actually spammers? A model of high precision conveys that it has a low false positive ($fn$) rate. Precision does not specify how many real spammers users were predicted as belonging to the non- spammers class - false negatives ($fn$). Recall or sensitivity deduces the ratio of accurately predicted spammers to all samples in the actual class. Recall answers the question: Amongst all users who are truly spammers, how many did we label? Neither recall nor precision can be solely used to evaluate the model performance in this binary classification task. Hence, F-Score is used that captures the properties of both recall and precision in one harmonic mean measure that is commonly used in training imbalanced datasets [83].

4.3.2 Experimental setting

The incorporated machine learning models are implemented and their hyperparameters are tuned using a random search strategy as illustrated in Table 6. Cross-validation has been conducted using three different k-fold values, namely 3,5 and 10. We found out that when k=10 the models perform best on an unseen dataset. We have not presented the results of each fold in each model due to the size of the paper and this is expected since a greater value of k commonly decreases the bias of the incorporated technique [84]. In each split, the total observations (i.e., 4000 users) are randomly divided into two datasets, namely the training dataset (80% of the total samples) and the validation dataset (20% of the total samples).

*Table 6: Hyperparameter settings*

| *Hyperparameter* | *Description* | *Examined Values* (Optimal values are in bold format) |
|---|---|---|
| **Random Forest Tree (RFT)** | | |
| *max_depth* | The longest route between the root and the leaf of the tree. | {10 , **20**, 30, 40, 100} |
| *min_sample_split* | The minimum number of observations that are required in any given node in order to split it. | {2 , **4**, 6} |
| *max_terminal_nodes* | Restricts the growth of the tree. | **none** |
| *max_features* | Features to consider when observing for the best split. | {"**auto**", "sqrt", "log2"} |
| *max_samples* | The number of samples to draw from original dataset | **none** |
| *criterion* | On which attribute will be split. | {"**gini**", "entropy"}, |
| *min_samples_leaf* | The minimum number of samples exist in the leaf node after splitting a certain node. | **5** |
| *n_estimators* | Specifies the number of trees in the forest of the model. | {**1**, 2, 3} |
| **Multi-Layer Feed-Forward Neural Network (MLFFNN)** | | |
| *activation* | The function used in the hidden layers . | {"**Rectifier**", "Tanh", "Maxout", "relu"} |
| *neurons/layer* | Number and size of the hidden layer. | {50, **100**, 200} |
| *#epochs* | Iteration times over dataset. | {**10**, 50} |
| *adaptive rate* | Unifies the benefits of momentum training and learning rate annealing | {0.1, **0.2**, 0.3} |
| *mean learning rate* | A non-negative scalar indicating step size | **0.003772** |
| *L1* | Regularization (absolute value of the weights) | **1.0E-5** |
| *L2* | Regularization (sum of the squared weights) | **0.0** |
| *Loss function* | loss (error) function | {"**Quadratic**", "CrossEntropy"} |
| **Gradient Boosted Classifier (GB)** | | |
| *#trees* | Number of generated trees. | {10, **20**, 30} |
| *loss* | loss function to be optimized. | {"**deviance**", "exponential"} |
| *learning_rate* | Assist in reducing the contribution of each tree. | **0.1** |
| *n_estimators* | The number of boosting stages to conduct. | {**100,** 200} |
| *criterion* | Measure the quality of tree split. | {"**friedman_mse**", "mse", "mae"} |
| *max threads* | Controls parallelism level of model building. | **1** |
| *max_depth* | Depth of the tree | **10** |
| **Generalised Linear Model (GLM)** | | |
| *family* | Uses binomial for classification. | {"gaussian", "**binomial**",} |
| *fit_intercept* | Indicates if bias or intercept should be added to the predictor. | {**True**, False} |
| *solver* | Used for optimisation. | {"**IRLSM**","L-BFGS"} |
| *Standardisation* | Standardisation numerical columns | **Checked** |
| *max number of threads* | Controls parallelism level of building model. | {**1**, 2, 3, 4} |
| **Decision Tree (DT)** | | |
| *criterion* | On which attribute will be split | {"**gain_ratio**", "information_gain"} |
| *max_depth* | Depth of the tree | {10, **20**} |
| *apply pruning* | Apply pruning after tree generation. | {**True**, False} |
| *confidence* | confidence level used for the pessimistic error calculation of pruning | **0.1** |
| *minimal gain* | The gain of a node is calculated before splitting it | 0.05 |
| **Naive Bayes (NB)** | | |
| *var_smoothing* | Slice of the variance of all features that is appended to variances for control stability. | **1e-9** |
| *laplace correction* | Prevents the occurrence of zero values. | **True** |
| *Priors* | Prior probabilities of the classes. | **None** |

4.3.3 Performance comparison of Machine Learning models

Figure 11 shows the ROC curves for each fold (k=10 folds) in each incorporated classifier. It provides an aggregate measure and shows the true positive rates against the false-positive rates at various threshold settings. The Area Under the ROC curve - AUC demonstrates the probability that a certain classifier ranks a randomly chosen positive observation (spammer) higher than a randomly chosen negative one (non- spammer). As illustrated in Figure 11, MLFFNN , RFT, and GBT achieve the best ROC means among all other classification models. MLFFNN embodies a sophisticated underlying structure that can learn and make intelligent decisions on its own. RF is also known as it is less overfit and robust on prediction [85] and performs well on complex and nonlinear data [86]. GLM is usually flexible and capable of performing unconventional analysis and is often utilised to analyse categorical forecaster variables [87]. On the other hand, NB classifier shows relatively worse performance compared to other incorporated models. This is mainly due to some internal assumptions that might lead NB model to perform inadequately. NB commonly presumes the independence of features and is not able to cope with the interactions of these features.

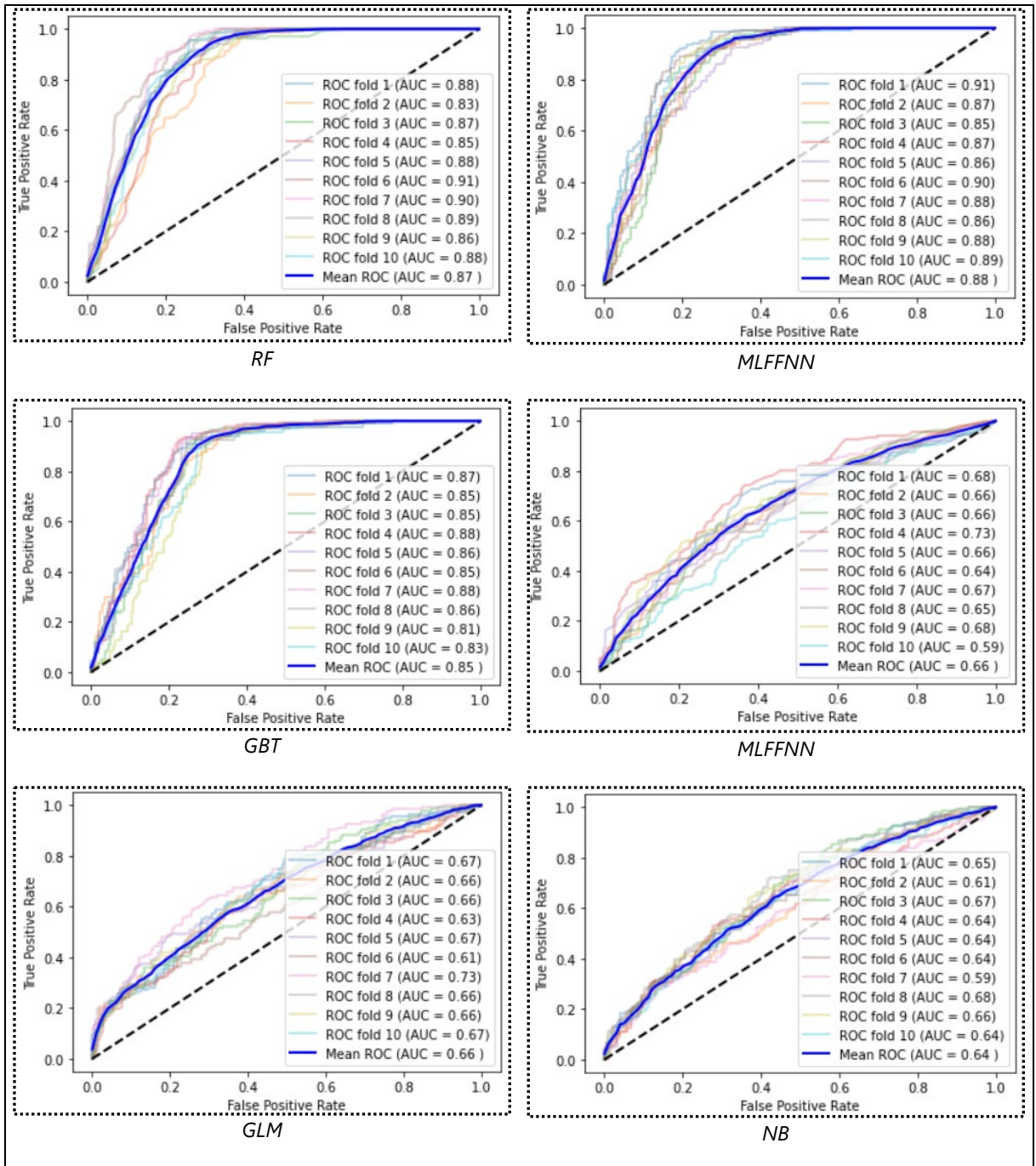

*Figure 11: ROC curves of all folds for incorporated machine learning algorithms*

**Error! Reference source not found.**Figure 12 illustrates the performance comparison between all implemented classifiers respectively. It can be seen that despite some convergence on the performance of some of the models, MLFFNN performs better in the classification task of this study; around 14% of the dataset used for validating the experiment were classified incorrectly by MLFFNN. However, some other models such as NB algorithm for example wrongly classified more samples in the prediction validations.

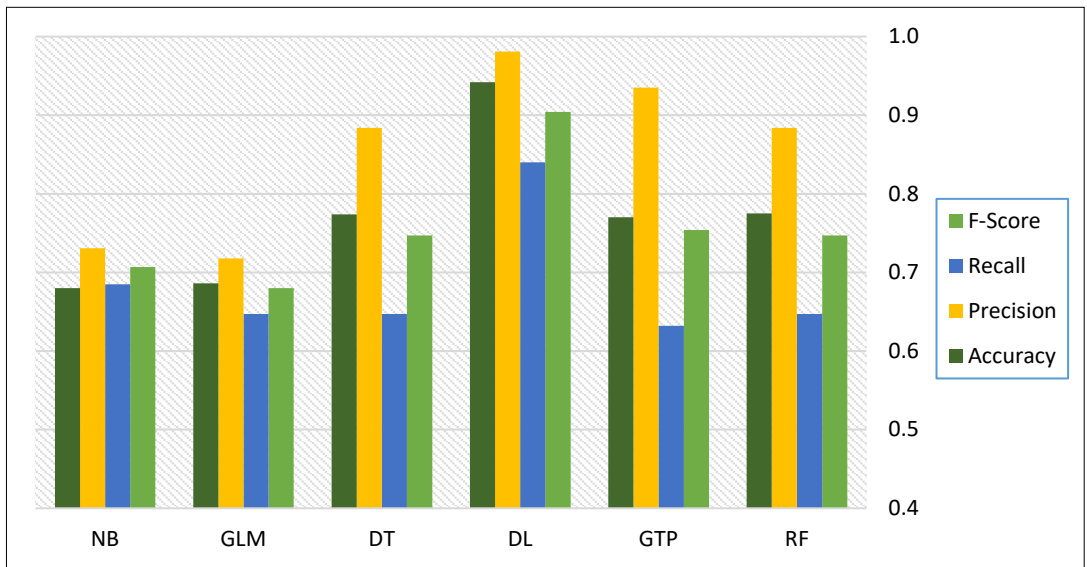

*Figure 12: Performance comparison of six classifiers to predict multi-topic spammers*

Figure 13 shows the highest estimated coefficient values computed for each feature using MLFFNN model. It shows that "topic frequency, $x_6$" is the highest estimated coefficient. This indicates that $x_6$ has the highest impact when compared with the other features. This is due to the importance of this feature in distinguishing the topics of interest. In particular, users who demonstrate an interest in a broad range of topics or few numbers of topics do not commonly convey legitimate behaviour as discussed in section 3.3.1.

| Attribute | Weight |
|---|---|
| x6 | 0.274 |
| x10 | 0.252 |
| x13 | 0.172 |
| x8 | 0.159 |
| x15 | 0.158 |
| x11 | 0.146 |
| x7 | 0.127 |
| x9 | 0.082 |
| x2 | 0.079 |
| x4 | 0.066 |
| x12 | 0.044 |
| x1 | 0.026 |
| x5 | 0.024 |
| x14 | 0.023 |
| x17 | 0.019 |
| x3 | 0.015 |
| x18 | 0.011 |
| x16 | 0.008 |

*Figure 13: Highest Positive Coefficients*

4.3.4 Social spam detection models – a baseline comparison

We undertake benchmark comparison with other spam detection approaches as well as with different versions of the proposed model over the curated labelled dataset. The list of models used in this comparison as well as the set of users retrieved from implementing each model are discussed as follows:

- ***Topic-dependent features (DDF):*** This method selects the set of users who attain the lowest and highest values in topic frequency ($x_6$) and Inverse topic frequency ($x_7$) as well as users who obtain the lowest value in all other topic-dependent features as indicated in Table 2.
- ***Topic-independent features (DIF)***: In this method, we implement a generic version of the proposed model. Therefore, we extract features of users with no consideration to users' topics of interest (i.e. extracted features are all topic-independent). The set of users retrieved based on this method are those who attain the lowest values in all extracted features ($x_6$ and $x_7$ were excluded).
- ***Low in-degree of topically anomalous (low in-degree TA) and social honeypot datasets (low in-degree SH)***: The high number of followers, referred to *high in-degree*, was used in previous studies to indicate influential users as well as spammers [88, 89]. We conduct this comparison on retrieved users who obtain the lowest number of followers (*low in-degree*) from the two selected datasets discussed in section 3.1.1.
- ***Don't Follow Me (DFM)*** [88]: this is one of the well-known attempts to detect social spammers using both content-based features and graph-based features. The model incorporates a number of machine learning classifiers such as Decision Tree, Neural Networks, SVM, and Naïve Bayesian. The model proposed in [88] is implemented and the set of users who achieved the lowest values in all features are retrieved for comparison.

**Evaluation Metric:** To report on the performance of each method, and due to the huge size of the dataset, we incorporate $AveragePrecision@k (AP@k)$. This measure is used as a weighted mean of precisions ($P@k$) achieved at each arbitrary threshold $k$. $P@k$ is computed based on how many spammers are present in the top-$k$ retrieved users in each method. The average precision evaluation metric is computed as follows:

$$APrecision@k = \frac{1}{N(k)} \sum_{i=1}^{k} (P@k \times rel@k)$$

Where $tp$ refers to the true positives, $N(k)$ is the total number of ground truth positives, and $rel@k$ is an indicator function that equals '1' if the user at rank $k$ is a spammer, zero otherwise.

**Model Comparison:** Figure 14**Error! Reference source not found.** shows the retrieval precision of the top-$k$ at 10, 20, 30, 40, 50 and the $APrecision@k$. As depicted in the figure, the evaluation results on the retrieved users of the proposed multi-topic social spammer detection verify the effectiveness of the developed approach to detect social spammer. This is evident as the topic-dependent features ($DDF$) method overshadows other incorporated and proposed baseline models. For example, the first ten users retrieved by enquiring users who obtained low and high topic frequency ($DDF$) were all exhibiting spamming behaviour. However, only three users of the first 10 retrieved users from the topically anomalous dataset, and whose in-degree features are the lowest, are spammers. Despite the good performance of the topic-independent features (DIF) method, along with don't follow me (DFM) approach, these approaches do not tackle the user's topic of interest, thus, the features are computed in general. However, the average precision accumulated using the proposed approach is promising for building spamming detection frameworks consolidated with the topic-based features proposed in this study.

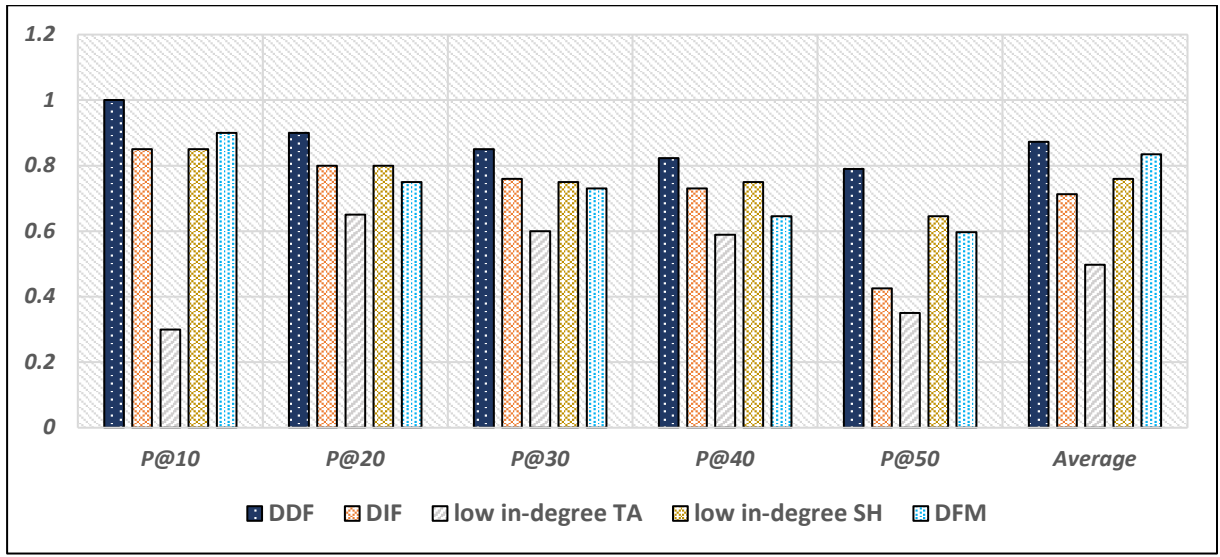

*Figure 14: Models comparison performance*

To add further clarification to the conducted benchmark comparison, we selected three true positive social spammers as identified by our model. Features of these users were extracted and the classification of these users based on other spammers classification approaches was inferred. As depicted in Table 7, low-in-degree models were unable to positively classify the users as spammers. This is evident as the idea of low-in-degree to identify spammers is built merely upon the followers-to-friends ratio ($x_{15}$) which is not always an adequate mechanism. Legitimate users might also convey an interest in following other users while attaining a low number of friends. DFM model [88] is still capable to identify spammers although the followers-to-friends ratio is one of its core features. This is because the original algorithm incorporates other graph and content-based features which offer it further consolidation, yet it is inadequate to detect spammers with other sophisticated features. Although the generic version of our model (i.e., DIF) demonstrates the ability to capture two out of three spammers, its generic nature incapacitates it from detecting users who publish contents in various topics ($x_6$ and $x_7$). On the other hand, our multi-topic social spammer detection model (i.e., DDF) is designed to be sensitive to capture such a spamming behaviour. Spammers attempt to gain credibility by establishing conversations addressing various topics, thereby persuading legitimate users to befriend them, or by hashtag hijacking or bulk messaging a certain topic or a variety of topics [1, 4]. By incorporating the novel distinguishing measure, semantic analysis, sentiment analysis, and other fine-grained features, our model overshadows other existing approaches. For example, although Spammer3 has a relatively balanced followers-to-friends ratio, other features such as topic frequency ($x_6$), Inverse topic frequency ($x_7$), topic-based #retweet ($x_8$), topic-based #likes ($x_9$), sum of the positive replies ($x_{11}$), and sum of the negative replies ($x_{12}$) conveys a suspicious behaviour.

*Table 7: Features extracted for three identified social spammers.*

| | | Spammer1 | Spammer2 | Spammer3 |
|---|---|---|---|---|
| **Features and their values** | $x_1$ | 18,861 | 12,767 | 19,914 |
| | $x_2$ | 538 | 766 | 238 |
| | $x_3$ | 101 | 62 | 212 |
| | $x_4$ | 34 | 21 | 83 |
| | $x_5$ | 21 | 20 | 56 |
| | $x_6$ | 5 | 3 | 8 |
| | $x_7$ | 0.699 | 0.47 | 0 |
| | $x_8$ | 16 | 11 | 42 |
| | $x_9$ | 66 | 128 | 7 |
| | $x_{10}$ | 45 | 95 | 250 |
| | $x_{11}$ | 6.947 | 11.821 | 5.76 |

|  |  |  |  |  |  |
|---|---|---|---|---|---|
|  | $x_{12}$ | -1.013 | -11.53 | -45.52 |
|  | $x_{13}$ | 910 | 952 | 581 |
|  | $x_{14}$ | 253 | 767 | 417 |
|  | $x_{15}$ | 0.073 | 0.013 | 0.028 |
|  | $x_{16}$ | 335 | 133 | 720 |
|  | $x_{17}$ | 201 | 97 | 408 |
|  | $x_{18}$ | 118 | 43 | 304 |
| Models Prediction | DDF | Spammer | Spammer | Spammer |
|  | DIF | Spammer | Spammer | Legitimate |
|  | low in-degree TA | Legitimate | Legitimate | Legitimate |
|  | low in-degree SH | Legitimate | Legitimate | Legitimate |
|  | DFM | Spammer | Spammer | Legitimate |

# 5 Discussion

In this paper, we have proposed a novel social spammer detection model for microblogging. We mainly focused on topic-dependent and topic-independent users' behaviours. We then conducted our experiments on the labelled twitter datasets using six popular machine learning algorithms (Random Forest Tree, Multi-Layer Feed-Forward Neural Network, Gradient Boosted Classifier, Generalised Linear Model, Decision Tree, and Naive Bayes). According to our experimental results, MLFFNN and RF approaches achieved the best ROC. The main reason behind the performance of MLFFNN in this experiment is the advantages that neural network-based approaches have compared to the traditional machine learning algorithms. They have their feature engineering capability and they focus on more relevant features for their predictions. Further, the implemented and trained RF model demonstrates an ability to capture spammers in the incorporated dataset. The good performance of the RF model is due to its adaptability and flexibility allowing it to search for the best features. RF operates on multiple and uncorrelated decision trees and the prediction of the model is based on the aggregated prediction values obtained by all trees which further consolidates the RF performance [90]. On the other hand, our analysis illustrates that NB has a poor prediction quality for this application. This is due to NB's too simplistic assumption that all features are independent.

Although our experimental results demonstrate good performance for our proposed approach in this study, next we need to improve it as follows:

- We will incorporate the shared textual contents by users in OSNs. There are a great set of features that we may extract from users' tweets/reviews/posts, e.g., their personality, emotional state, level of trustworthiness, and gender.
- We will design a deep neural-based classifier to act as a spam detection approach. We plan to use Graph Convolutional Network (GCN), because it allows converting the data of users and their features to a graph-based environment, where users are the nodes, and their features are the nodes' features.
- We will conduct further investigation on the users' credibility and trustworthiness level and incorporate them in our social spamming detection analysis. This will be attained by proposing context-aware social spammer detection model considering geographical location and temporal factors.

# 6 Conclusion

With the coronavirus COVID-19 pandemic continuing, the risk of spreading rumours, misinformation and spam through social media dominates. On the other side of the spectrum, efforts pursue to spot and stop spammers and their activities over the social networks. This paper attempts to consolidate these endeavours by implementing an effective and robust approach to detect social spammers. With a distinctive vector of features constructed by performing fine-grained topic-based analysis of users' contents, various machine learning algorithms are implemented and trained over a labelled dataset. The performance of these algorithms is evaluated using important evaluation indicators. Further, the model is benchmarked against existing techniques used to detect social spammers and anomalous users. The model overshadows the baseline models which verifies the utility of the proposed model in the designated task.

In future work, we will incorporate the temporal dimension. Users' behaviours in OSNs may vary over time. It follows that their credibility changes as well; hence, the temporal factor should be considered. Further, a new topic-based graph model will be designed and implemented to proliferate the users' credibility throughout the entire social network. Therefore, we aim to study the structure of online social networks and consider integrating semantics of the textual content and the temporal factor.